\documentclass[12pt]{article}
\pdfoutput=1

\usepackage[a4paper,text={16.8cm,22.4cm}]{geometry}
\usepackage{amsmath,amsfonts,braket,slashed,amssymb,tikz,bm,psfrag,graphicx,color,dsfont}
\usepackage{multicol}
\usepackage{ctable}
\usepackage[small,labelfont=bf]{caption}
\usepackage{hyperref}

\RequirePackage[sort&compress,square,comma,numbers]{natbib}
\allowdisplaybreaks
\RequirePackage[sort&compress,square,comma,numbers]{natbib}
\allowdisplaybreaks
\begin{document}

\begin{titlepage}

\begin{flushright}
\normalsize
MITP/18-075\\
CERN-TH-2018-199\\
August 29, 2018

\end{flushright}

\vspace{1.0cm}
\begin{center}
\Large\bf\boldmath
Axion-Like Particles at Future Colliders
\end{center}

\vspace{0.5cm}
\begin{center}
Martin Bauer$^a$, Mathias Heiles$^{b}$, Matthias Neubert$^{b,c}$ and Andrea Thamm$^d$\\
\vspace{0.7cm} 
{\sl ${}^a$Institut f\"ur Theoretische Physik, Universit\"at Heidelberg\\
Philosophenweg 16, 69120 Heidelberg, Germany\\[3mm]
${}^b$PRISMA Cluster of Excellence \& Mainz Institute for Theoretical Physics\\
Johannes Gutenberg University, 55099 Mainz, Germany\\[3mm]
${}^c$Department of Physics \& LEPP, Cornell University, Ithaca, NY 14853, U.S.A.\\[3mm]
${}^d$Theoretical Physics Department, CERN, 1211 Geneva, Switzerland}\\
\end{center}

\vspace{0.8cm}
\begin{abstract}

Axion-like particles (ALPs) are pseudo Nambu-Goldstone bosons of spontaneously broken global symmetries in high-energy extensions of the Standard Model (SM). This makes them 
a prime target for future experiments aiming to discover new physics which addresses some of the open questions of the SM. While future high-precision experiments 
can discover ALPs with masses well below the GeV scale, heavier ALPs can be searched for at future high-energy lepton and hadron colliders. We discuss the 
reach of the different proposed colliders, focusing on resonant ALP production, ALP production in the decay of heavy SM resonances, and associate ALP production with photons, $Z$ bosons or Higgs bosons. We consider the leading effective operators mediating interactions between the ALP and SM 
particles and discuss search strategies for ALPs decaying promptly as well as ALPs with delayed decays. Projections for the high-luminosity run of the LHC and its high-energy upgrade, CLIC, the future $e^+e^-$ ring-colliders CEPC and FCC-ee, the future $pp$ colliders SPPC and FCC-hh, and for the MATHUSLA surface array are presented. We further discuss the constraining power of future measurements of electroweak precision parameters on the relevant ALP couplings.  
\end{abstract}

\end{titlepage}

\section{Introduction}

Axion-like particles (ALPs) are light, gauge-singlet pseudoscalar particles with derivative couplings to the Standard Model (SM). The name is inspired by the QCD axion, which is the pseudo-Nambu-Goldstone boson associated with the breaking of the Peccei-Quinn symmetry \cite{Peccei:1977hh, Peccei:1977ur, Weinberg:1977ma, Wilczek:1977pj}, proposed to address the strong CP problem.
More generally, ALPs appear in any theory with a spontaneously broken global symmetry and possible ALP masses and couplings to SM particles range over many orders of magnitude. In certain regions of parameter space ALPs can be non-thermal candidates for Dark Matter \cite{Preskill:1982cy} or, in other regions where they decay, mediators to a dark sector. For large symmetry breaking scales, the ALP can be a harbinger of a new physics sector at a scale $\Lambda$ which would otherwise be experimentally inaccessible. Since the leading ALP couplings to SM particles scale as $\Lambda^{-1}$, ALPs become weakly coupled for large new-physics scales. Accessing the smallest possible couplings is thus crucial to reveal non-trivial information about a whole new physics sector. 

Depending on the region in parameter space spanned by the ALP mass and couplings, the search strategies 
vary greatly. For masses below twice the electron mass, the ALP can only decay into photons and the corresponding decay rate scales like the third power of the ALP mass. Thus, light ALPs are usually long-lived and travel long distances before decaying.  Experiments probing long-lived ALPs include helioscopes such as CAST \cite{Arik:2008mq}, SUMICO \cite{Inoue:2008zp,Graham:2015ouw}, as well as observations from the evolution of red giant stars \cite{Raffelt:1985nk,Raffelt:1987yu,Raffelt:2006cw} and the Supernova SN1987a \cite{Payez:2014xsa,Jaeckel:2017tud}. In addition, a set of cosmological constraints from the modification to big-bang nucleosynthesis, distortions of the cosmic microwave background and extragalactic background light measurements exclude a large region of this parameter space and are sensitive to very small ALP-photon couplings \cite{Cadamuro:2011fd,Millea:2015qra}. For intermediate ALP masses up to the GeV scale, collider experiments such as BaBar, CLEO, LEP and the LHC searching for missing-energy signals probe long-lived ALPs with non-negligible couplings to SM particles \cite{Balest:1994ch,delAmoSanchez:2010ac}. Current and future beam-dump searches are sensitive to ALPs with masses below $\sim1\,$GeV radiated off photons and decaying outside the target \cite{Riordan:1987aw, Bjorken:1988as,Alekhin:2015byh,Dobrich:2015jyk}. ALP couplings to other SM particles are generally less constrained than the ALP-photon coupling. ALP couplings to charged leptons are constrained by searches for ALPs produced in the sun \cite{Armengaud:2013rta}, the evolution of red giants \cite{Raffelt:2006cw}, by beam-dump experiments \cite{Essig:2010gu}, and through associate ALP production at BaBar \cite{TheBABAR:2016rlg, Bauer:2017ris}. Proposals for future experiments suggest measuring the ALP-electron coupling in Compton scattering of an electron in the background of low- and high-intensity electromagnetic fields \cite{Dillon:2018ypt,Dillon:2018ouq}.

High-energy colliders are sensitive to a large and previously inaccessible region in parameter space \cite{Bauer:2017nlg,Bauer:2017ris}. 
Requiring the ALP to decay within the detector opens up a new region of parameter space.  The different ALP production mechanisms at colliders offer a rich phenomenology, allowing us to probe a large range of ALP masses and couplings. 
Beyond resonant production, ALPs can be produced in decays of heavy SM particles \cite{Dobrescu:2000jt,Dobrescu:2000yn,Chang:2006bw,Draper:2012xt,Curtin:2013fra,Bauer:2017nlg,Bauer:2017ris} or in association with gauge bosons, Higgs bosons or jets \cite{Mimasu:2014nea, Jaeckel:2015jla, Knapen:2016moh, Brivio:2017ije}. Resonant ALP production is particularly powerful for small new-physics scales $\Lambda$, because the production rate is proportional to $1/\Lambda^2$. ALP production in Higgs and $Z$ decays, on the other hand, is sensitive to large new-physics scales $\Lambda$, because the corresponding exotic Higgs or $Z$ branching fractions are enhanced by the small widths of these bosons. Interesting channels at the LHC are the on-shell decays $h\to a a$, $h\to Z a$ and $Z\to \gamma a$. Dedicated analyses by the LHC experiments will provide new and complementary ALP searches. ALPs can also be produced in the decay of B mesons \cite{Frere:1981cc, Hiller:2004ii, Batell:2009jf, Freytsis:2009ct, Andreas:2010ms, Dolan:2014ska, Izaguirre:2016dfi}. These decays are sensitive to flavor-changing ALP couplings, which we will not consider in this work. In an upcoming publication we will discuss constraints from flavor-changing ALP couplings including ALPs produced in the decay of B mesons \cite{flavorpaper}.

Depending on the ALP mass and coupling structure, ALPs produced at colliders can decay into photons, charged leptons, light hadrons or jets.  These decays can be prompt or displaced if the width of the ALP is sufficiently small. We present bounds from current and future high-energy collider searches
for ALPs decaying into photons, charged leptons and jets, including the case where the ALP couples dominantly to gluons. Existing constraints on the ALP-gluon coupling come from mono-jet \cite{Mimasu:2014nea} and di-jet \cite{ATLAS:2016bvn} searches at the LHC and the rare kaon decay $K^+\to \pi^+ a$ mediated by ALP-pion mixing \cite{Fukuda:2015ana}.

Future hadron colliders can operate at unprecedented center-of-mass energies, whereas future lepton 
colliders benefit from their clean collision environment and the large production rates of on-shell $Z$ bosons and tagged Higgs bosons. Two current proposals for circular electron-positron colliders are the Circular Electron-Positron Collider (CEPC) based in China \cite{CEPC-SPPCStudyGroup:2015csa} and the $e^+ e^-$ Future Circular 
Collider (FCC-ee) based at CERN \cite{FCCeedata}. CEPC is envisioned to have a $50\,$km tunnel and operate both at the $Z$ pole and as a Higgs factory (at $\sqrt{s} = 250\,$GeV). At the $Z$ pole the target is to produce $10^{10}$ $Z$ bosons per year. Over a period of $10$ years an integrated luminosity of $5\,$ab$^{-1}$ should be accumulated at two interaction points, which 
corresponds to one million Higgs events \cite{CEPC-SPPCStudyGroup:2015csa}.
The FCC-ee is a proposed ring collider with 80\,--\,100\,km circumference operating at center-of-mass energies between $90$ and $400\,$GeV. At the FCC-ee, more than $10^{12}$  $Z$ bosons would be produced at four interaction points within one year \cite{SlidesJanot}. Roughly three million Higgs bosons would be produced in five years. 
Linear lepton colliders such as the ILC or CLIC loose in luminosity compared to their circular counterparts. The ILC is proposed to operate at $250, 350$ or $500\,$GeV, accumulating an integrated luminosity of $2, 0.2$ and $4\,$ab$^{-1}$, respectively \cite{Baer:2013cma,Fujii:2017vwa}. CLIC is designed to collect $0.5, 1.5$ and $3\,$ab$^{-1}$ at $380\,$GeV, $1500\,$GeV and $3\,$TeV center-of-mass energy, respectively \cite{CLIC:2016zwp}. 

Current proposals for high-energy proton colliders include the High-Energy LHC (HE-LHC) operating at 27\,TeV in the existing LHC tunnel and accumulating $15\,$ab$^{-1}$ \cite{HELHC}, the FCC-hh based at the proposed CERN FCC-ee tunnel operating at a 
center-of-mass energy of $100\,$TeV with a target luminosity in the range of 10\,--\,20\,ab$^{-1}$ per experiment \cite{Hinchliffe:2015qma}, and the Super-Proton-Proton-Collider (SPPC) based in the CEPC tunnel in China operating at 70\,--\,100\,TeV \cite{CEPC-SPPCStudyGroup:2015csa} accumulating 3\, ab$^{-1}$. 

Comparing the regions of ALP parameter space that can be probed with these future hadron and lepton colliders is particularly interesting and contributes to corroborating the physics case for these various machines. In this work we also consider proposed new experiments searching for long-lived particles, such as FASER \cite{Feng:2017uoz}, Codex-B \cite{Gligorov:2017nwh} and MATHUSLA \cite{Chou:2016lxi}, which can access the ALP parameter space 
between the regions covered by LHC experiments and bounds from cosmology.

This paper is structured as follows: In Section~\ref{sec:ALPBackground} we review the effective Lagrangian for an ALP interacting with SM fields and introduce the formalism for our ALP detection strategy. In Section~\ref{sec:futurecolliders} we discuss the reach of ALP searches at future colliders. We focus on existing LEP and LHC limits in Section~\ref{sec:LEPLHC},  ALP searches at lepton colliders in Section~\ref{sec:leptoncolliders}, and move on to ALP searches at hadron colliders in Section~\ref{sec:hadcol}. In Section~\ref{sec:MATHUSLA} we discuss the reach of the future surface detector MATHUSLA at the LHC. Section~\ref{sec:conclusions} contains our conclusions.

%
\section{ALP production and decays}
\label{sec:ALPBackground}
%

%
\subsection{Effective Lagrangian}
%

An ALP is a light scalar which is a singlet under the SM gauge group and odd under CP. The ALP Lagrangian respects a shift symmetry, which is only softly broken by a mass term. Its leading interactions with the SM particles are described by dimension-5 operators \cite{Georgi:1986df} 
\begin{equation}\label{Leff}
\begin{aligned}
   {\cal L}_{\rm eff}
   &= \frac12 \left( \partial_\mu a\right)\!\left( \partial^\mu a\right)
    - \frac{m_{a}^2}{2}\,a^2 
    + \sum_f \frac{c_{ff}}{2}\,\frac{\partial^\mu a}{\Lambda}\,\bar f\gamma_\mu\gamma_5 f \\[-1mm]
   &\quad\mbox{}+ g_s^2\,C_{GG}\,\frac{a}{\Lambda}\,G_{\mu\nu}^A\,\tilde G^{\mu\nu,A}
    + g^2\,C_{WW}\,\frac{a}{\Lambda}\,W_{\mu\nu}^A\,\tilde W^{\mu\nu,A}
    + g^{\prime\,2}\,C_{BB}\,\frac{a}{\Lambda}\,B_{\mu\nu}\,\tilde B^{\mu\nu} \,,
\end{aligned}
\end{equation}
where the couplings to fermions $c_{ff}$ are assumed to be flavor universal, and $\Lambda$ sets the characteristic scale of global symmetry breaking. The commonly used axion decay constant $f_a$ is related to our new-physics scale by $\Lambda/|C_{GG}^{\rm eff}|=32\pi^2 f_a$. ALPs can obtain part of their mass from non-perturbative dynamics but need additional explicit breaking of the shift symmetry to be heavier than the QCD axion.\footnote{Models in which the SM gauge symmetry is extended can also lead to larger ALP masses \cite{Rubakov:1997vp, Berezhiani:2000gh, Fukuda:2015ana, Belyaev:2016ftv, Cacciapaglia:2017iws, Agrawal:2017ksf, Bellazzini:2017neg}. } In the absence of an explicit breaking term, the QCD axion is defined by a strict relation between its 
mass and decay constant, $m_a \propto f_\pi m_\pi/f_a$, with $f_\pi $ and $m_\pi$ the pion decay constant and mass, respectively. For ALPs such a strict relation does not apply, since $m_a$ and $f_a$ are independent parameters. 

In the broken phase of the electroweak symmetry, the ALP couples to the photon and the $Z$ boson as
\begin{equation}\label{gammaZ}
   {\cal L}_{\rm eff}
   \ni e^2\,C_{\gamma\gamma}\,\frac{a}{\Lambda}\,F_{\mu\nu}\,\tilde F^{\mu\nu}
    + \frac{2e^2}{s_w c_w}\,C_{\gamma Z}\,\frac{a}{\Lambda}\,F_{\mu\nu}\,\tilde Z^{\mu\nu}
    + \frac{e^2}{s_w^2 c_w^2}\,C_{ZZ}\,\frac{a}{\Lambda}\,Z_{\mu\nu}\,\tilde Z^{\mu\nu} \,.
\end{equation}
The relevant Wilson coefficients are given by
\begin{equation}\label{WilsonCoeff}
C_{\gamma\gamma}=C_{WW}+C_{BB}, \hspace{1cm} 
C_{\gamma Z}=c^2_w\,C_{WW}-s^2_w\,C_{BB}, \hspace{1cm} 
C_{ZZ}= c^4_w\,C_{WW}+ s^4_w\,C_{BB}\,,
\end{equation}
where $s_w$ and $c_w$ are the sine and cosine of the weak mixing angle, respectively. The exotic decay $Z\to\gamma a$ is governed by the Wilson coefficient $C_{\gamma Z}$. 

Note that the anomaly equation for the divergence of the axial-vector current allows us to rewrite the ALP-fermion couplings in \eqref{Leff} in the form
\begin{equation}
   \frac{c_{ff}}{2}\,\frac{\partial^\mu a}{\Lambda}\,\bar f\gamma_\mu\gamma_5 f 
   = - c_{ff}\,\frac{m_f}{\Lambda}\,a\,\bar f\,i\gamma_5 f 
    + c_{ff}\,\frac{N_c^f Q_f^2}{16\pi^2}\,\frac{a}{\Lambda}\,e^2 F_{\mu\nu}\,\tilde F^{\mu\nu} + \dots \,,
\end{equation}  
where the dots represent similar terms involving gluons and weak gauge fields \cite{Bauer:2017ris}. This is instructive to relate results obtained for the ALP with analogous, and maybe more familiar, results derived for a CP-odd Higgs boson. E.g.~the first term on the right-hand side is now of the same form as the coupling of a CP-odd Higgs to fermions.

Interactions with the Higgs boson, $\phi$, appear only at dimension-6 and higher, 
\begin{equation}\label{LeffD>5}
   {\cal L}_{\rm eff}^{D\ge 6}
   = \frac{C_{ah}}{\Lambda^2} \left( \partial_\mu a\right)\!\left( \partial^\mu a\right) \phi^\dagger\phi
    + \frac{C_{Zh}}{\Lambda^3} \left( \partial^\mu a\right) 
    \left( \phi^\dagger\,iD_\mu\,\phi + \mbox{h.c.} \right) \phi^\dagger\phi + \dots \,,
\end{equation}
where the first operator mediates the decay $h\to aa$, while the second one is responsible for $h\to Za$. Note that a possible dimension-5 operator coupling the 
ALP to the Higgs current vanishes by the equations of motion. However, in theories where a heavy new particle 
acquires most of its mass through electroweak symmetry breaking, the non-polynomial dimension-5 operator 
\begin{equation}\label{Leffnonpol} 
\frac{C_{Zh}^{(5)}}{\Lambda} \left( \partial^\mu a\right)
    \left( \phi^\dagger\,iD_\mu\,\phi + \mbox{h.c.} \right) \ln\frac{\phi^\dagger\phi}{\mu^2} 
\end{equation}
can be present \cite{Bauer:2016ydr,Bauer:2016zfj,Bauer:2017nlg,Bauer:2017ris}. In our analysis we allow for the presence of such an operator.

We now summarise the relevant partial widths needed for the remainder of this paper. We express the relevant decay rates in terms of effective Wilson coefficients, which take into account loop-induced contributions, that have been calculated in \cite{Bauer:2017ris}. In the case of $h \to Z a$ decay the effective coefficient is defined as $C_{Zh}^{\rm eff} = C_{Zh}^{(5)} + C_{Zh} v^2/2 \Lambda^2 + \text{loop effects}$. The relevant ALP decay rates are
\begin{align}
 \Gamma(a\to\gamma\gamma)  &= \frac{4\pi\alpha^2 m_a^3}{\Lambda^2}\,\big| C_{\gamma\gamma}^\text{eff} \big|^2 \,, \\
 \Gamma(a\to \ell^+ \ell^-)&=\frac{m_a m_\ell^2}{8\pi\Lambda^2} \left| c_{\ell\ell}^\text{eff}\right|^2\sqrt{1-\frac{4m_\ell^2}{m_a^2}}\,
 ,\\
\Gamma(a\to gg) &= \frac{32\pi\,\alpha_s^2(m_a)\,m_a^3}{\Lambda^2}
    \left[ 1 + \frac{83}{4}\,\frac{\alpha_s(m_a)}{\pi} \right] \left| C_{GG}^\text{eff} \right|^2\,,
\end{align}
where the latter expression is only valid if $m_a \gg \Lambda_\text{QCD}$. The exotic Higgs and $Z$-boson decay rates into ALPs are given by  
\begin{align}
\Gamma(h \to Za)&=\frac{m_h^3}{16\pi\,\Lambda^2}|C_{Zh}^\text{eff}|^2\lambda^{3/2}\Big(\frac{m_Z^2}{m_h^2},\frac{m_a^2}{m_h^2}\Big)\,,\\
\Gamma(h \to aa)&=\frac{m_h^3\,v^2}{32\pi\,\Lambda^4}|C_{ah}^{\rm eff}|^2\left(1-\frac{2m_a^2}{m_h^2}\right)^2\sqrt{1-\frac{4m_a^2}{m_h^2}}\,,\\
\Gamma(Z\to \gamma a)&=\frac{8\pi\alpha\,\alpha(m_Z)\,m_Z^3}{3s_w^2 c_w^2\Lambda^2}\,\big| C_{\gamma Z}^\text{eff} \big|^2
    \left( 1 - \frac{m_a^2}{m_Z^2} \right)^3,
\end{align}
where $\lambda(x,y)=(1-x-y)^2-4xy$.

%
\subsection{ALP production at colliders} 
%

At high-energy colliders, ALPs can be produced in different processes. We distinguish resonant production through gluon or photon fusion and $e^+e^-$ annihilation, the production in
association with photons, $Z$ bosons, Higgs bosons or jets  \cite{Mimasu:2014nea, Jaeckel:2015jla, Knapen:2016moh, Brivio:2017ije}, and the production via exotic decays of on-shell Higgs or $Z$ bosons \cite{Bauer:2017nlg, Bauer:2017ris}.

\subsubsection*{Resonantly produced ALPs}

At high-energy colliders, ALPs can be produced resonantly through gluon-fusion $gg\to a$ (GGF), photon fusion $\gamma\gamma \to a$ ($\gamma\gamma$F), or electron-positron annihilation $e^+e^-\to a$. If an ALP coupling to heavy gauge bosons is present, ALPs can also be produced in vector-boson fusion \cite{Buttazzo:2018qqp}. An important difference between resonant production and ALP production through exotic decays or associated ALP production 
is that the resonant production cross section is always suppressed by the ALP mass, $m_a$, over the new physics scale $\Lambda$. Resonant production is 
therefore mostly relevant for large ALP masses. At hadron colliders large ALP masses are also important to suppress backgrounds. The cross sections for the 
resonant ALP production processes are 
\begin{align}
\sigma_\text{GGF}(pp \to a)&=\frac{4\pi^3 \alpha_s^2(m_a)}{s}\frac{m_a^2}{\Lambda^2} |C_{GG}^\text{eff}|^2\,K_{a\to gg}f\hspace{-.18cm}f_{gg} \left(\frac{m_a^2}
{s} \right)\,, \\
\sigma_{\gamma\gamma \text{F}}(pp \to a)&=\frac{\pi^3 \alpha^2(m_a)}{2s}\frac{m_a^2}{\Lambda^2} |C_{\gamma\gamma}^\text{eff}|^2\, f\hspace{-.18cm}
f_{\gamma\gamma} \left(\frac{m_a^2}{s} \right)\,, \\
\sigma(e^+ e^- \to a)&\stackrel{s\, \approx\, m_a^2}{=}\frac{4\pi \Gamma_a}{(s-m_a^2)^2+m_a^2\Gamma_a^2}\frac{\sqrt{s} m_e^2}{8\pi \Lambda^2}|c_{ee}^\text{eff}|^2
\end{align}
where $f\hspace{-.18cm}f_{gg} (y) = \int_y^1 \frac{dx}{x}f_{g/p}(x)f_{g/p}(y/x)$ is the gluon luminosity function (the photon luminosity function is defined 
analogously) and $K_{a\to gg}\approx 3.3 - 2.4$ for $m_a=100-1000\,$GeV accounts for higher-order QCD corrections \cite{Harlander:2002wh, Ahmed:2016otz}. In the last equation we set $m_e^2/s\to 0$. 
Both $\sigma(e^+e^-\to a)$ as well as the quark contribution to $\sigma(pp\to a)$ are strongly suppressed by the light fermion masses and these processes are therefore not the dominant production modes. 
ALP production in photon fusion with a subsequent di-photon decay of the ALP is particularly interesting, because the production times decay rate only depends on the ALP mass and the single coupling $C_{\gamma\gamma}^{\rm eff}$. Furthermore, the uncertainty of the photon distribution function in the proton has recently been considerably improved allowing for more robust limits \cite{Manohar:2016nzj}.
For resonantly produced ALPs finite-lifetime effects do not play any role because the sizeable couplings and ALP masses required to obtain appreciable production cross sections lead to prompt ALP decays. 

%
\begin{figure}
\begin{center}
\includegraphics[width=0.63\textwidth]{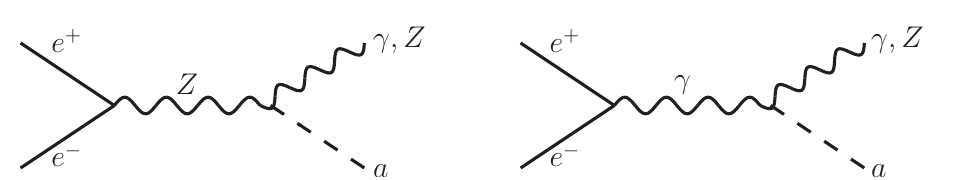}
\includegraphics[width=0.3\textwidth]{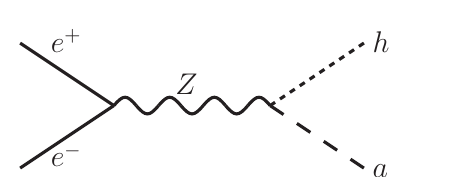}
\end{center}
\vspace{-2mm}
\caption{\label{fig:eegaa} Tree-level Feynman diagrams for the processes $e^+ e^-\to X a$ with $X=\gamma, Z, h$.}
\end{figure}
%

\subsubsection*{\boldmath ALP production in association with a photon, $Z$ or Higgs boson}
An important production mechanism especially at $e^+e^-$ colliders is associated ALP production. The relevant Feynman diagrams are shown in Figure~\ref{fig:eegaa}. Additional diagrams 
with ALPs radiated off an initial-state electron  are suppressed by $m_e^2/s$ relative to the shown graphs and hence neglected here. ALPs can be radiated of a photon or a $Z$ boson and thereby be produced in association with a $\gamma$, a $Z$ or a Higgs.  The differential cross sections for ALPs  produced in association with a $\gamma$, a $Z$ or a Higgs boson are given by 
\begin{align}\label{sigmahighE1}
   \frac{d \sigma(e^+ e^-\to\gamma a)}{d\Omega} =
   & \, 2 \pi \alpha \alpha^2(s) \frac{s^2}{\Lambda^2} \left(1 - \frac{m_a^2}{s}\right)^3 \left(1 + \cos^2 \theta \right) \left(|V_\gamma(s)|^2 + |A_\gamma(s)|^2\right), \\ \vspace{0.1cm} \label{sigmahighE2}
     \frac{d \sigma(e^+ e^-\to Z a)}{d \Omega} =
    & \, 2 \pi \alpha \alpha^2(s) \frac{s^2}{\Lambda^2} \, \lambda^{\frac{3}{2}} \left(x_a, x_Z \right)\left(1 + \cos^2 \theta \right) \left(|V_Z(s)|^2 
+ |A_Z(s)|^2\right), \\ \vspace{0.1cm} \label{sigmahighE3}
      \frac{d \sigma(e^+ e^-\to h a)}{d\Omega} 
   = & \, \frac{\alpha}{128\pi \, c_w^2 s_w^2} \frac{|C_{Zh}^{\rm eff}|^2}{\Lambda^2}  \frac{s \, m_Z^2}{(s-m_Z^2 )^2}\, \lambda^{\frac{3}{2}} \left(x_a,x_h \right) \sin^2 \theta \, (g_V^2 + g_A^2) \,,
\end{align}
where $x_i =m_i^2/s$ and 
\begin{align}\label{sigmahighEb}
V_\gamma(s) &= \frac{C_{\gamma \gamma}^\text{eff}}{s} + \frac{g_V}{2 c_w^2 s_w^2}\frac{C_{\gamma Z}^\text{eff}}{s - m_Z^2+ i m_Z \Gamma_Z}\, , \hspace{2cm} A_\gamma(s) = 
\frac{g_A}{2 c_w^2 s_w^2}\frac{C_{\gamma Z}^\text{eff}}{s - m_Z^2+ i m_Z \Gamma_Z} \,, \\ \vspace{0.1cm}
V_Z(s) &= \frac{1}{c_w s_w}\frac{C_{\gamma Z}^\text{eff}}{s} + \frac{g_V}{2 c_w^3 s_w^3}\frac{C_{Z Z}^\text{eff}}{s - m_Z^2+ i m_Z \Gamma_Z} \,,  \hspace{1cm} A_Z(s)= \frac{g_A}{2 
c_w^3 s_w^3}\frac{C_{Z Z}^\text{eff}}{s - m_Z^2+ i m_Z \Gamma_Z}\,,
\end{align}
and $g_V = 2 s_w^2 - 1/2$ and $g_A=-1/2$. Note that the cross sections with a gauge boson in the final state become independent of $s$ in the high-energy limit $m_a^2, m_Z^2 \ll s 
< \Lambda$, while the cross section for $e^+ e^-\to h a$ decreases as $1/s$ in this limit.  

Light or weakly coupled ALPs can be long-lived, and thus only a fraction of them decays inside the detector and can be reconstructed. The average ALP decay length perpendicular to the beam axis is given by
\begin{equation}\label{eq:Lperp}
   L_a^\perp(\theta) = \frac{\sqrt{\gamma_a^2 - 1}}{\Gamma_a}\,\sin\theta \,, 
\end{equation}
where $\Gamma_a$ denotes the total width of the ALP, $\theta$ is the scattering angle (in the center-of-mass frame) and $\gamma_a$ specifies the relativistic boost factor. For the case of associated ALP production with a boson $X = \gamma, Z, h$, we have
\begin{align}
\gamma_a=\frac{s-m_X^2+m_a^2}{2m_a\sqrt{s}}.
\end{align}
In order to obtain the total cross sections for ALPs produced in associated production, we integrate the differential distributions \eqref{sigmahighE1}\,--\,\eqref{sigmahighE3} with the non-decay probability, i.e.
\begin{align}\label{eq:epemdecay}
\sigma(e^+e^-\to X a)=\int d\Omega\, \frac{d\sigma(e^+e^-\to a X)}{d\Omega} \left(1-e^{-L_\text{det}/L_a^\perp (\theta)}\right)\,,
\end{align}
where $L_\text{det}$ is the transverse distance from the beam axis to the detector component relevant to the reconstruction of the ALP. 

Associated production at hadron colliders will not be considered here. For long-lived or invisibly decaying ALPs such processes have been explored recently in \cite{Mimasu:2014nea, Brivio:2017ije}.

\subsubsection*{\boldmath ALP production in exotic decays of on-shell Higgs and $Z$ bosons }
\begin{figure}
\includegraphics[width=\textwidth]{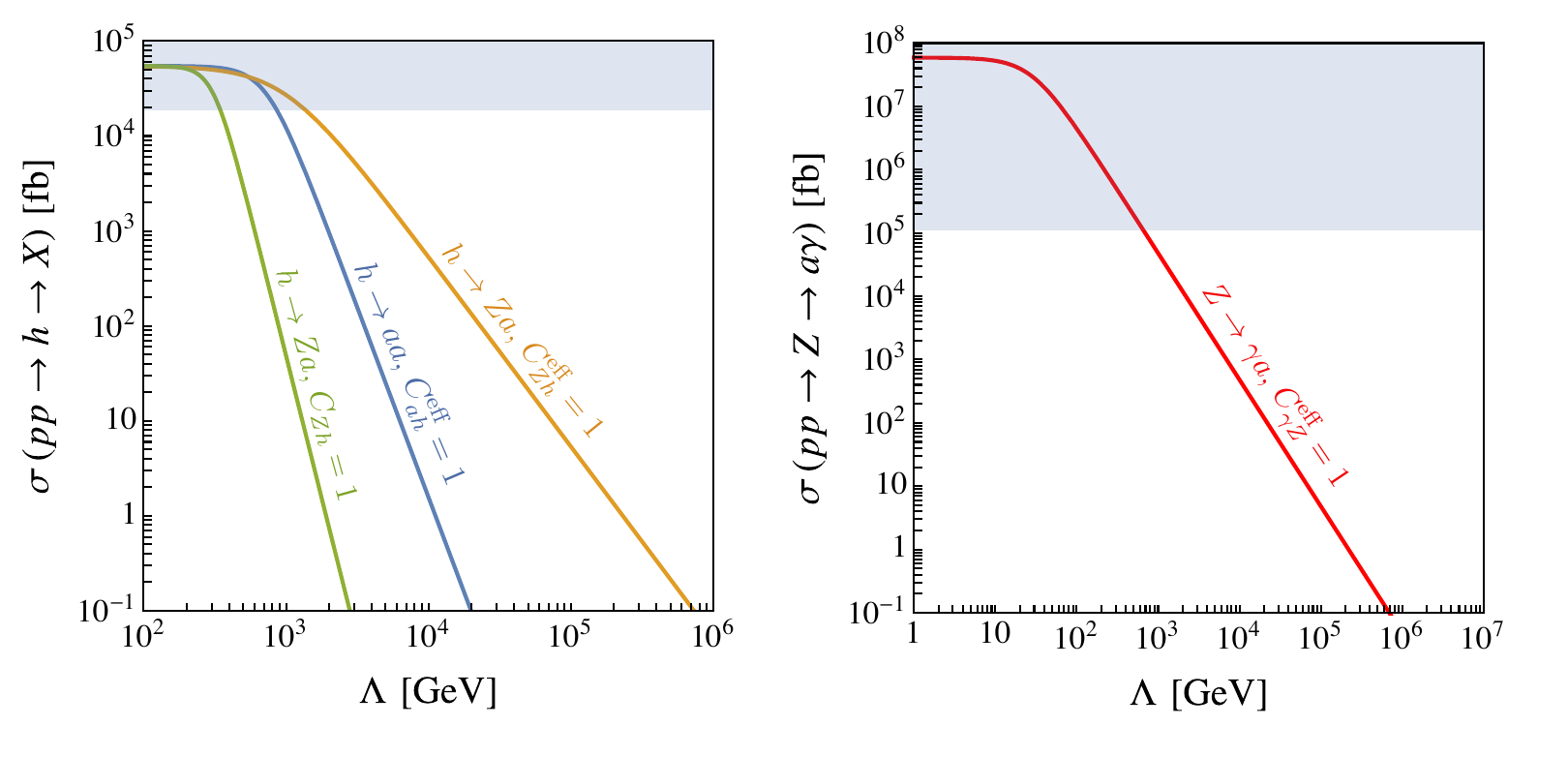}
\caption{\label{fig:ALPpsec} Production cross sections of ALPs produced in the decays of Higgs and $Z$ bosons at the LHC ($\sqrt{s} = 14\,$TeV) versus the new-physics scale $\Lambda$. We set $m_a=0$ and fix the relevant Wilson coefficients to 1. For the green contour in the left plot, we fix $C_{Zh}^{(5)}=0$ and only consider the dimension-7 coupling in \eqref{LeffD>5}.
The grey regions in the two plots are excluded by Higgs coupling measurements and the measurement of the total $Z$ width, respectively. }
\end{figure}

Exotic decays are particularly interesting, because even small couplings can lead to appreciable branching ratios. In the case of the Higgs boson, the SM decay widths are strongly suppressed, and consequently the  branching ratios for Higgs decays into ALPs can be as large as several percent \cite{Bauer:2017nlg, Bauer:2017ris}. In the case of the $Z$ boson, the huge samples of $Z$ events expected at future colliders provide sensitivity to $Z \to \gamma a$ branching ratios much below current bounds.  
This allows us to probe large new-physics scales $\Lambda$, as illustrated in Figure \ref{fig:ALPpsec}, where we show the cross sections of the processes $pp \to Z \to \gamma a$, $pp \to h \to Z a$ and $pp \to h \to aa$  at the LHC with $\sqrt{s} = 14\,$TeV. The figure nicely reflects the different scalings of the dimension-5, 6, and 7 operators in the effective ALP Lagrangian. The 
shaded region in the left plot is excluded by Higgs coupling measurements constraining general beyond the SM decays of the Higgs boson, $\text{Br}(h\to\text{BSM})<0.34$ 
\cite{Khachatryan:2016vau}. The shaded area in the right plot is derived from the measurement of the total $Z$ width, which corresponds to $\text{Br}(Z\to \text{BSM})<0.0018$ \cite{ALEPH:2005ab}. This leads to constraints on the coefficients $|C_{Zh}^{\textrm{eff}}| < 0.72\,(\Lambda/\textrm{TeV})$, $|C_{ah}^{\rm eff}| < 1.34\,(\Lambda/\textrm{TeV})^2$ and $|C_{\gamma Z}^{\rm eff}| < 1.48\,(\Lambda/\textrm{TeV})$. 
The Higgs and $Z$-boson production cross sections at $14\,$TeV are given by $\sigma(pp\to h)= 54.61\,$pb  \cite{Anastasiou:2016cez} and $\sigma(pp\to Z)=60.59$ nb, computed at NNLO using tools provided in \cite{Grazzini:2017mhc,Hamberg:1990np}.

%
%

 As discussed above, it is important to include the effects of a possible finite ALP decay length. Using the fact that most Higgs and $Z$ bosons are produced in the forward direction at the LHC and approximating the ATLAS and CMS detectors (as well as future detectors) by infinitely long cylindrical tubes, we first perform a Lorentz boost to the rest frame of the decaying boson. In this frame the relevant boost factor for the Higgs or $Z$ decay into ALPs are given by
\begin{align}
\gamma_a=\begin{cases}\displaystyle{ \frac{m_h^2-m_Z^2+m_a^2}{2m_am_h}}\,;& \text{for}\,\, h \to Z a\,,\\[14pt]
\displaystyle \frac{m_h}{2m_a}\,;&\text{for}\,\, h \to aa\,,\\[14pt]
\displaystyle\frac{m_a^2+m_Z^2}{2m_Zm_a}\,;&\text{for}\,\, Z \to \gamma a\,.\\
\end{cases}
\end{align}
We can compute the fraction of ALPs decaying before they have travelled a certain distance $L_{\rm det}$ from the beam axis, finding 
\begin{equation}\label{eq:ffactors}
\begin{aligned}
   f_{\rm dec}^{a} &= \int_0^{\pi/2}\!d\theta \,  \sin\theta
    \left( 1 - e^{-L_{\rm det}/L_a^\perp(\theta)} \right) \,, \\
   f_{\rm dec}^{aa} &= \int_0^{\pi/2}\!d\theta \,  \sin\theta
    \left( 1 - e^{-L_{\rm det}/L_a^\perp(\theta)} \right)^2 \,,
\end{aligned}
\end{equation} 
where $f^a_\text{dec}$ is relevant for $h \to Z a$ and $Z\to \gamma a$ decays, and $f^{aa}_\text{dec}$ applies to $h\to aa$ decays. 

%
\begin{figure}
\centering
\includegraphics[width=0.45\textwidth]{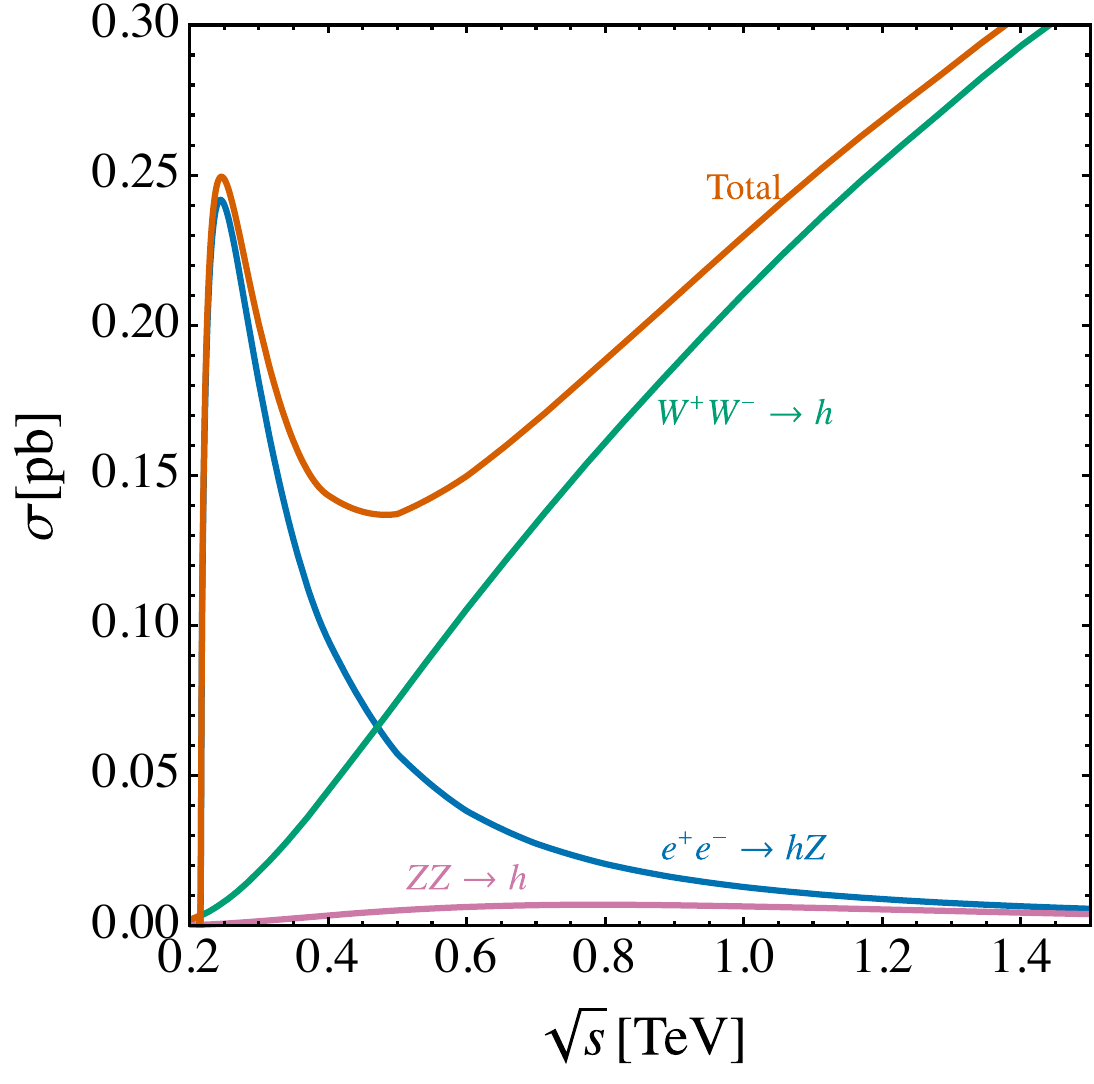}
\caption{\label{fig:production} Leading order Higgs production cross sections at $e^+e^-$ colliders 
as a function of the center-of-mass energy, produced with \texttt{MadGraph5} \cite{Alwall:2014hca}. }
\end{figure}
%
%

For Higgs bosons produced at $e^+e^-$ colliders the assumption of forward production is no longer justified. Rather, the angular distribution in the scattering angle $\vartheta$ of the Higgs boson in the center-of-mass frame are approximately given by \cite{Miller:2001bi}
\begin{align}
\frac{d\sigma}{d\Omega}&\propto\begin{cases}\dfrac{3}{2}\dfrac{\lambda(x_h,x_Z)\sin^2\vartheta+8x_Z}{\lambda(x_h,x_Z)
+12x_Z}\,~\stackrel{s\gg m_h^2}{\longrightarrow} ~\dfrac{3}{2}\sin^2\vartheta\,; &\quad e^+e^-\to h Z\,,\\[5mm]
\dfrac{3}{2}\sin^2\vartheta\,; &\quad \quad \text{VBF}\,,\\
\end{cases}
\end{align}
with $x_i =m_i^2/s$. The approximation $s \gg m_h^2 $ for the Vector Boson Fusion (VBF) process is justified, because the VBF cross section becomes the dominant production cross section for $\sqrt{s}\gtrsim 500\,$GeV \cite{Barger:1993wt, Miller:2001bi}. This fact is illustrated in Figure~\ref{fig:production}, which depicts the cross section of various Higgs production modes at lepton colliders as functions of the center-of-mass energy. Even though in the Higgs rest frame, the angular distribution of the produced ALPs will be isotropic, the corresponding distribution in the center-of-mass frame is more complicated in this case. Since the Higgs bosons are predominantly produced with $\vartheta\approx 90^\circ$, we will for simplicity make the conservative assumption that the ALPs are also produced at maximum scattering angle in the center-of-mass frame, corresponding to $\sin\theta=1$ in \eqref{eq:Lperp}. For the resonant process $e^+e^-\to Z\to \gamma a$ on the $Z$ pole, no such difficulty arises. The corresponding differential branching ratio can be obtained from \eqref{sigmahighE1} by setting $s=m_Z^2$, and the decay-length effect can be taken into account as shown in \eqref{eq:epemdecay}.

For prompt ALP decays, we demand all final state particles to be detected in order to reconstruct the decaying SM particle. For the decay into photons we 
require the ALP to decay before the electromagnetic calorimeter which, at ATLAS and CMS, is situated approximately $1.5\,$m from the interaction point, and we thus take $L_{\rm det} = 
1.5\,$m. Analogously, the ALP should decay before the inner tracker, $L_{\rm det} = 2\,$cm, for an $e^+e^-$ final state to be detected. We also require $L_\text{det}
=2\,$cm for muon and tau final states in order to take full advantage of the tracker information in reconstructing these events. For CLIC, we use $L_\text{det}=0.6\,$m for lepton reconstruction \cite{CLICdp-Note-2017-001}. We define the effective branching ratios
\begin{align}
\text{Br}(h\to Za\to Y\bar{Y}+X\bar{X})\big\vert_\text{eff}&=\text{Br}(h\to Za)\,\text{Br}(a\to X\bar{X})f_\text{dec}^a\,\text{Br}(Z\to Y\bar{Y})\,,\label{eq:LHChZa}\\
\text{Br}(h\to aa\to X\bar{X}+X\bar{X})\big\vert_\text{eff}&=\text{Br}(h\to aa)\,\text{Br}(a\to X\bar{X})^2f_\text{dec}^{aa}\,,\label{eq:LHChaa} \\
\text{Br}(Z\to \gamma a \to \gamma X\bar{X} )\big\vert_\text{eff}&=\text{Br}(Z\to \gamma a)\,\text{Br}(a\to X\bar{X})f_\text{dec}^a\,,\label{eq:LHCZag}
\end{align}
where $X=\gamma, e, \mu, \tau, \text{jet}$ and $Y=\ell, {\rm hadrons}$. Multiplying the effective branching ratios by the appropriate Higgs or $Z$ production cross sections and luminosity allows us to 
derive results for a specific collider.  At hadron colliders like the LHC, we require $100$ signal events, since this is what is typically needed to suppress backgrounds in new-physics searches with prompt decays of Higgs and $Z$ bosons \cite{Khachatryan:2016vau, Chatrchyan:2013vaa,Aad:2015bua} (see also \cite{Bauer:2017ris} for 
further discussion). At lepton colliders we assume a much cleaner environment and show the reach for 4 signal events. 

We do not take advantage of the additional background reduction obtained by cutting on a secondary vertex in the case where the ALP lifetime becomes appreciable. A dedicated analysis by the experimental collaborations including detailed simulations of the backgrounds is required to improve on our projections.

%
\begin{table} \centering
\begin{tabular}{ c | c c c c c c}
  	 	& Collisions 	& $ \sqrt{s}$ [TeV]	& $L$ [ab$^{-1}$] 	& \# $Z$ bosons 	& \# Higgs bosons & References \\ \hline		
  ILC$_{250}$		& $e^+ e^-$ 	& $0.25$ 	& $2$			& $\sim 2 \times 10^7$ & $\sim 500 \times 10^3$ & \cite{Fujii:2017vwa}\\
   ILC$_{350}$		& $e^+ e^-$ 	& $0.35$ 	& $0.2$			& $\sim 9 \times 10^5$ & $\sim 30 \times 10^3$ & \cite{Fujii:2017vwa}\\
      ILC$_{500}$		& $e^+ e^-$ 	& $0.5$ 	& $4$		& $\sim 9 \times 10^6$ & $\sim 550 \times 10^3$ & \cite{Fujii:2017vwa}\\
  CLIC$_{380}$ 	& $e^+ e^-$ 	& $0.38$ 	& $0.5$			& $\sim 2 \times 10^6$ & $89\times 10^3$ & \cite{CLIC:2016zwp} \\
    CLIC$_{1500}$ 	& $e^+ e^-$ 	& $1.5$ 	& $1.5$			& $\sim 4 \times 10^5$ & $420\times 10^3$ & \cite{CLIC:2016zwp} \\
  CLIC$_{3000}$ 	& $e^+ e^-$ 	& $3$ 	& $3$			& $\sim 2 \times 10^5$ & $926\times 10^3$ & \cite{CLIC:2016zwp} \\
  CEPC	& $e^+ e^-$ 	& $0.091$ 		& $0.1$			& $10^{10}$	& &  \cite{CEPC-SPPCStudyGroup:2015csa} \\
    CEPC 	& $e^+ e^-$ 	& $0.25$ 		& $5$				&  	& $10^6$ &  \cite{CEPC-SPPCStudyGroup:2015csa} \\
  FCC-ee	& $e^+ e^-$ 	& $0.091$ 			& $145$			&  $10^{12}$	& & \cite{FCCeedata}\\
		& $e^+ e^-$ 	& $0.161$ 			& $20$			& 	&  $10^6$ & \cite{FCCeedata}\\
  		&	$e^+ e^-$ 	        & $0.25$ 			& $5$			& 	
				        & $10^6$ 
				        & \cite{FCCeedata}\\  \hline
  LHC 	& $pp$ 		& $14$ 			& $3$			& & &\\
  HE-LHC 	& $pp$ 		& $27$ 			& $15$			& & & \cite{HELHC} \\
  SPPC  	& $pp$ 		& $100$ 			& $3$			& & & \cite{CEPC-SPPCStudyGroup:2015csa}\\
  FCC-hh 	& $pp$ 		& $100$ 			& $20$			& & & \cite{Hinchliffe:2015qma}\\ 
\end{tabular} \caption{\label{tab:benchmarks}Benchmark specifications of various future collider proposals. The number of $Z$ and Higgs bosons indicated with a $\sim$ have been computed with \texttt{MadGraph5} \cite{Alwall:2014hca}. }
\vspace{-.2cm}
\end{table}
%

%
\section{Collider reach for ALP searches}
\label{sec:futurecolliders}
%
The reach of ALP searches at current and future colliders depends on the type of collider, the ALP production mechanism, and the center-of-mass 
energy of the experiment. For the LHC and the most advanced proposals for future colliders, we use the benchmark specifications collected in Table~\ref{tab:benchmarks}.  In the following, we determine the reach of future colliders in comparison to the high-luminosity phase of the LHC with $\sqrt{s}=14\,$TeV and 
an integrated luminosity of $L=3\,$ab$^{-1}$.

%
\begin{figure}[t]
\begin{center}
\includegraphics[width=1\textwidth]{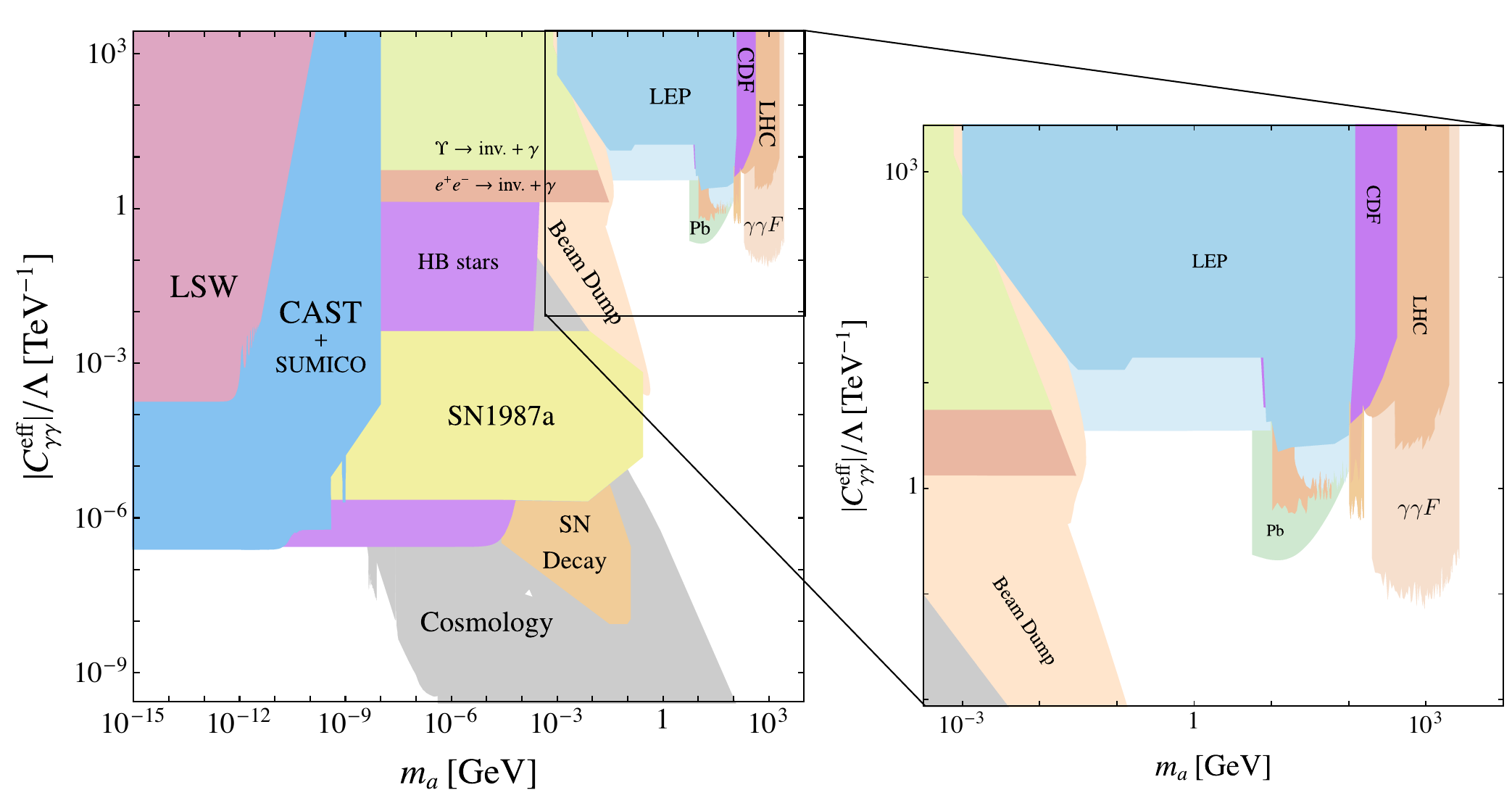}
\end{center}
\vspace{-3mm}
\caption{\label{fig:Cggbounds} Left: Summary plot of constraints on the parameter space spanned by the ALP mass and ALP-photon coupling. Right: Enlarged display of the constraints from collider searches: LEP (light blue and blue), CDF (purple), LHC from associated production and $Z$ decays (orange), LHC from photon fusion (light orange), and from heavy-ion collisions at the LHC (green). }
\end{figure}
%

%
\begin{figure}[t]
\begin{center}
\includegraphics[width=0.495\textwidth]{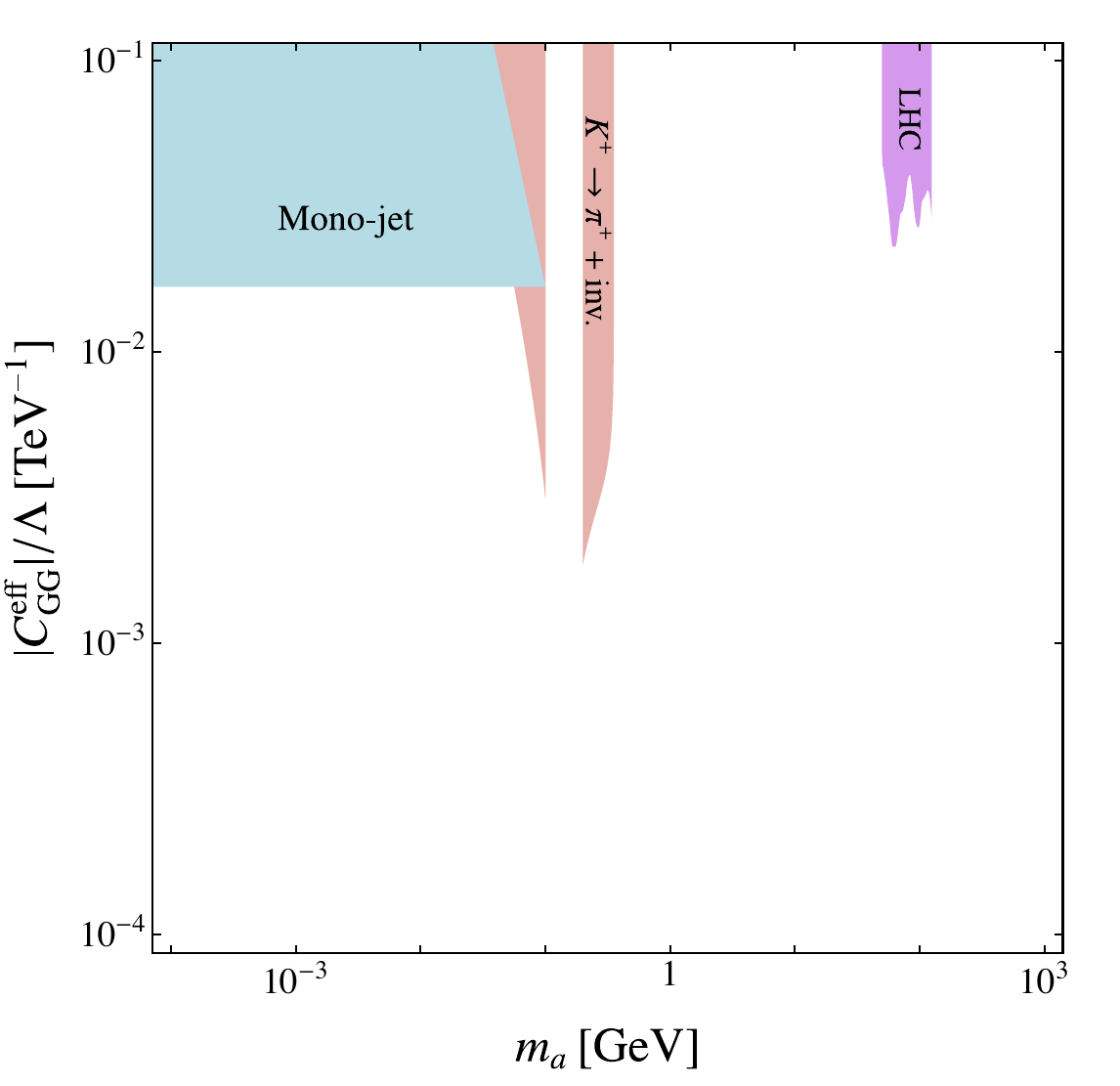}
\includegraphics[width=0.48\textwidth]{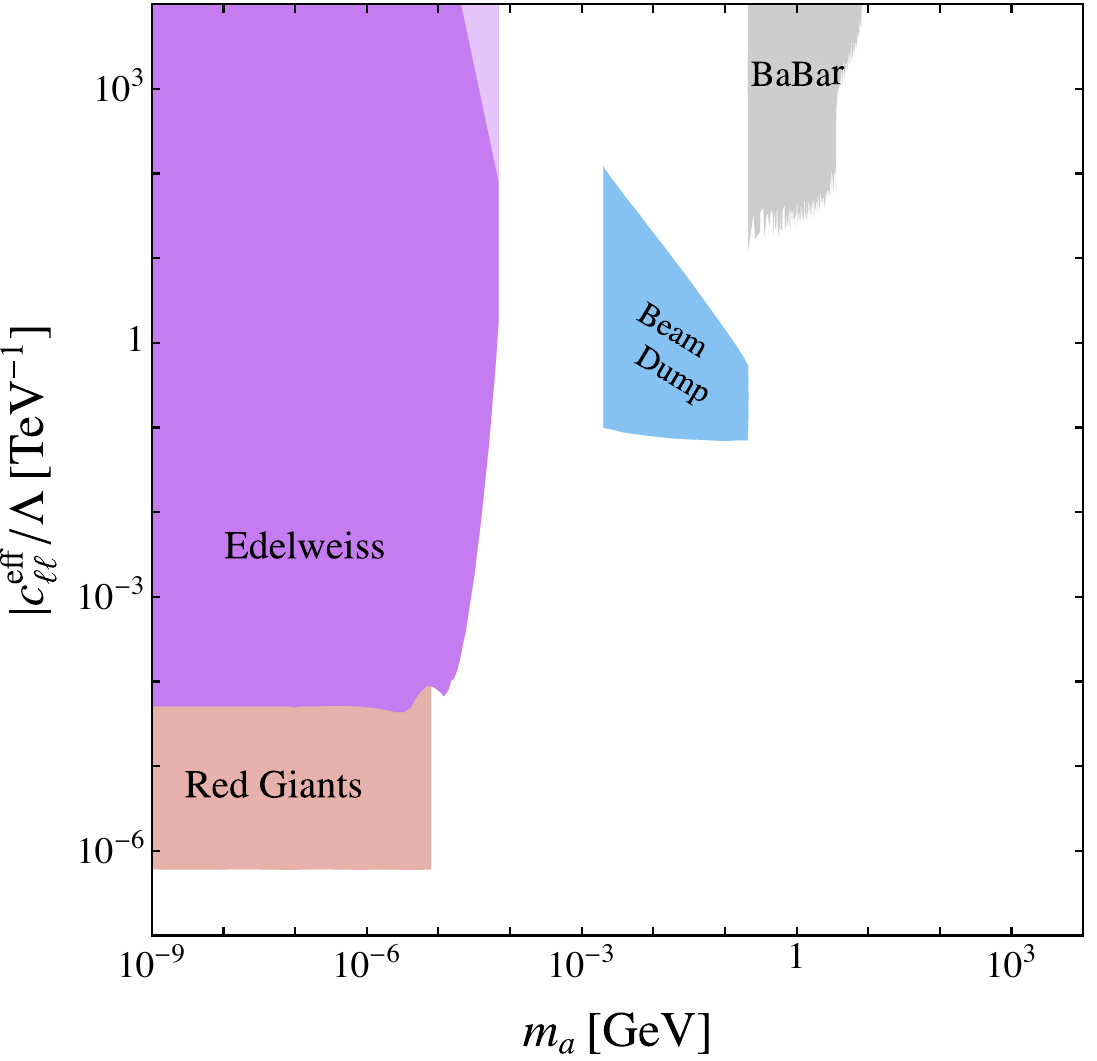}
\end{center}
\vspace{-3mm}
\caption{\label{fig:Cllbounds}  Left: Existing constraints on the ALP mass and coupling to gluons by mono-jet searches at the LHC (light blue), rare kaon decays (light red) and three-jet events (purple). Right:  Constraints on the ALP mass and coupling to leptons from searches for solar axions (purple), the evolution of red giants (light red), beam dump searches for ALP decays into electrons (blue) and BaBar searches for $e^+e^-\to 4 \mu$. }
\end{figure}
%

\subsection{ALP searches at the LHC and LEP}
\label{sec:LEPLHC}

Constraints from ALP searches at LEP have been discussed for the associated production of ALPs with a photon and the subsequent ALP decay into photon 
pairs ($e^+e^-\to  \gamma a\to 3 \gamma$) \cite{Mimasu:2014nea}, as well as for on-shell $Z$ decays ($e^+e^- \to Z \to  \gamma a\to 3 \gamma$) \cite{Jaeckel:2015jla}. The excluded parameter space in the $m_a - |C_{\gamma\gamma}^{\rm eff}|/\Lambda$ plane is shown in blue in Figure~\ref{fig:Cggbounds}. 
At the LHC, exotic Higgs and $Z$ boson decays are the most promising search channels. Decays of on shell $Z$ bosons at the LHC have been discussed in 
\cite{Jaeckel:2015jla, Mimasu:2014nea, Brivio:2017ije, Bauer:2017ris}. The constraints from these searches can be mapped onto 
the $m_a - |C_{\gamma\gamma}^{\rm eff}|/\Lambda$ plane under the assumption that the two couplings $C_{\gamma\gamma}^\text{eff}$ and $C_{\gamma Z}^\text{eff}$ are related to each other. For example, if the ALP couples to hypercharge but not to $SU(2)_L$, then \eqref{WilsonCoeff} implies $C_{\gamma Z}=-s_w^2\,C_{\gamma\gamma}$, since $C_{WW}=0$. The corresponding constraint is shown in orange in Figure~\ref{fig:Cggbounds}.\footnote{The LHC reach is slightly enhanced for the scenario 
$C_{BB}=0$, cf.\ Figure~23 in \cite{Bauer:2017ris}.} The purple region is excluded by Tevatron searches for $p\bar p \to 3 \gamma$ \cite{Aaltonen:2013mfa}, again assuming $C_{WW}=0$.

The dark green area in Figure~\ref{fig:rareZdecayshh} in Section~\ref{sec:hadcol} below depicts the region where 100 events are expected in the process $pp\to Z \to  \gamma a \to 3 \gamma$ at the LHC with $\sqrt{s}=14\,$TeV and $L=3\,$ab$^{-1}$. We demand that the ALPs decay before they reach the 
electromagnetic calorimeter $L_\text{det}=1.5\,$m. Note that for a part of this parameter space the photons from the ALP decay are very boosted and hard to 
distinguish from a single photon in the detector \cite{Toro:2012sv}.  
Searches for the exotic Higgs decays $pp\to h \to Z a\to Z\gamma\gamma$ and $pp\to h \to a a\to 4\gamma$ cannot be translated into constraints in the $m_a - |C_{\gamma\gamma}^{\rm eff}|/\Lambda$ plane, because the ALP-Higgs couplings governed by the coefficients $C_{Zh}^{\rm eff}$ and $C_{ah}^{\rm eff}$ are generally not related to $C_{\gamma\gamma}^{\rm eff}$. Instead, we 
show the reach of the high-luminosity LHC in the $|C_{Zh}^{\rm eff}|/\Lambda - |C_{\gamma\gamma}^{\rm eff}|/\Lambda$ or $|C_{ah}^{\rm eff}|/\Lambda^2 - |C_{\gamma\gamma}^{\rm eff}|/\Lambda$ 
planes for some fixed ALP masses in Figure~\ref{fig:pphZa} in Section~\ref{sec:hadcol}.

Besides ALP production in exotic decays of Higgs and $Z$ bosons, ALP production through photon fusion plays an important role at the LHC. This process 
was first considered in a VBF-type topology in \cite{Jaeckel:2012yz}, and the excluded region is part of the orange shaded region in Figure~\ref{fig:Cggbounds}. For 
GeV-scale ALPs produced in photon-fusion, (quasi-)elastic heavy-ion collisions can provide even stronger constraints due to the large charge of the lead ions ($Z=82$) 
used in the LHC heavy-ion collisions \cite{Knapen:2016moh, Knapen:2017ebd}. The parameter space probed by this process is shown in green in Figure~\ref{fig:Cggbounds}.

Recently, the parton distribution function of the photon has been determined with significantly improved accuracy \cite{Manohar:2016nzj}, and searches for di-photon 
resonances at the LHC can be recast to give bounds on heavy pseudoscalar particles with couplings to photons 
\cite{Molinaro:2017rpe}. We have computed the constraints based on the most recent ATLAS analysis with $39.6\,$fb$^{-1}$ of data \cite{Aaboud:2017yyg} 
and show the corresponding sensitivity regions in light orange in Figure~\ref{fig:Cggbounds}. A recent proposal to search for ALPs in elastic photon scattering at the LHC allows for a similar reach in the $m_a - |C_{\gamma\gamma}^{\rm eff}|/\Lambda$ plane \cite{Baldenegro:2018hng}.

Searches for ALPs decaying into photons are motivated by the relation between the ALP coupling to gluons $C_{GG}^{\rm eff}$ and to photons $C_{\gamma\gamma}^{\rm eff}$ in models addressing the strong CP problem, and from a practical point of view by the difficulty of observing light ALPs decaying into jets at hadron colliders. On the other hand, if the coupling to gluons is present in the effective ALP Lagrangian \eqref{Leff}, constraints arise from searches for mono-jets at ATLAS and CMS \cite{Mimasu:2014nea}, as well as from the rare kaon decay $K^+\to \pi^+ a$ mediated by ALP-pion mixing \cite{Fukuda:2015ana}.\footnote{We thank Yotam Soreq for pointing out an error in our calculation of the constraint derived from $K^+\to \pi^+ a$ decays. During the publication process of this paper a thorough analysis of bounds from ALP-gluon couplings has appeared \cite{Aloni:2018vki}.}  
Di-jet searches at the LHC can provide bounds on heavy ALPs with masses $m_a> 1\,$TeV, whereas recent searches for a new vector resonance decaying into di-jets accompanied by hard initial state radiation $pp \to  j Z'\to 3 j$ can be recast into limits on ALPs with masses below the TeV scale in the process $pp \to  j a\to 3 j$ \cite{ATLAS:2016bvn, Sirunyan:2017nvi, Aaboud:2018zba}.\footnote{Limits from di-jet searches from previous experiments sensitive to lower masses are weak \cite{Dobrescu:2013coa}, and we do not show them in Figure~\ref{fig:Cllbounds}.} As pointed out in \cite{Mariotti:2017vtv}, the hard cut on hadronic activity applied in the analyses \cite{ATLAS:2016bvn, Sirunyan:2017nvi, Aaboud:2018zba}, strongly reduces the efficiency in a gluon-fusion initiated signal compared to a $q\bar q$-initiated signal as expected for a vector resonance. In Figure~\ref{fig:Cllbounds}, we show the limit derived in \cite{Mariotti:2017vtv} (labeled LHC) in the  $m_a-|C_\text{GG}^\text{eff}|/\Lambda$ plane.  

Another promising signature are leptonically decaying ALPs: $a\to \ell^+\ell^-$ with $\ell=e,\mu,\tau$. In the right panel of Figure~\ref{fig:Cllbounds} we show a compilation of current limits in the $m_a-|c_{\ell\ell}^\text{eff}|/\Lambda$ plane taken from \cite{Bauer:2017ris}. We assume universal couplings to leptons, such that lepton flavor changing couplings mediated by ALP exchange are absent at tree level. Lepton colliders are sensitive to the resonant production of ALPs with subsequent decays into leptons. In general, however, the loop-induced couplings to $Z\gamma$ and $\gamma\gamma$ are more important than the tree-level coupling to electrons because the latter is suppressed by $m_e/\Lambda$. Even for ALPs coupling only to leptons at tree level the associated production cross sections via the processes shown in Figure~\ref{fig:eegaa} dominate over the $e^+e^-$ annihilation cross section. 
Projections for additional signatures, such as $pp \to a W^\pm (\gamma)$, $pp \to a jj (\gamma)$, $pp \to h a$ and $pp \to  t\bar t a$ with stable ALPs or invisible ALP decays have been considered in \cite{Brivio:2017ije}.  The complementarity between di-photon and di-lepton final states has also been emphasised in the proposal for boosted di-tau resonances \cite{Cacciapaglia:2017iws}.

\subsection{ALP searches at future lepton colliders}
\label{sec:leptoncolliders} 

Future lepton colliders have the potential to precisely measure the 
properties of the Higgs boson and search for new physics effects in electroweak observables. In addition they offer qualitatively new ways to search for ALPs. 
In contrast to hadron colliders, $e^+e^-$ machines offer a much cleaner detector environment allowing one to identify ALPs produced in association with a $Z$ boson, 
a photon or a Higgs boson. Therefore, in addition to ALPs produced in exotic decays of on-shell $Z$ and Higgs bosons, we also discuss the associated production of ALPs.\footnote{See \cite{Frugiuele:2018coc} for a study of these channels for the case of a relaxion.} On the contrary, barring a fine-tuning of the collider energy, the resonant production of ALPs cannot be observed in $e^+e^-$ collisions.\footnote{The radiative return process is suppressed by a factor $m_e^2/s$.}

Of particular interest are processes governed by a single non-vanishing Wilson coefficient at tree-level that allow us to compare the projected sensitivity reach of the future lepton colliders with the results of previous experiments, see Figures~\ref{fig:Cggbounds} and \ref{fig:Cllbounds}. Studying these processes at a lepton collider allows one in particular to probe benchmark models in which the ALP couples only to electroweak gauge bosons or only to charged leptons. Other processes involve different couplings for the production and the decay of the ALP. 
Among these, the rich Higgs program of all proposed future lepton colliders motivates the search for ALPs produced in association with Higgs bosons or in exotic Higgs decays. For these channels, in order to compare the reach of the various proposed experiments, we focus on the di-photon and di-lepton ALP decay channels. Following \cite{Bauer:2017ris}, we present the corresponding sensitivity regions in a two-dimensional plane spanned by these two couplings. We derive this sensitivity region by demanding 4 reconstructed events before the inner tracker for ALPs decaying into electrons and muons and before the ECAL for ALP decays into photons. The assumption that this number of events is sufficient for future lepton colliders to be sensitive to a signal is based on the very clean final states (photons or leptons) and the strong cuts that can be applied if several resonances appear in the signal, e.g.~the ALP, the $Z$ boson and the Higgs in the process $h \to  Z a $. For similar searches at LEP, cuts have reduced the background to 2-9 events \cite{Acciarri:1994gb, Anashkin:1999da, Mimasu:2014nea}. We emphasise that these projected sensitivity regions therefore represent estimates that cannot replace a full analysis that should be performed by experimentalists. Analogous studies could be performed for different ALP decay channels, such as $a\to b \bar b$ or $a \to jj$.

\subsubsection*{\boldmath ALP production in association with a photon, $Z$ or Higgs boson}

For $e^+e^-\to \gamma a\to 3 \gamma$ and $e^+e^-\to Z a\to Z\gamma \gamma$, the process only depends on the photon coupling $|
C_{\gamma\gamma}^{\rm eff}|/\Lambda$ once a specific relation between $C_{WW}$ and $C_{BB}$ is assumed, see \eqref{WilsonCoeff}. The projected reach can therefore be compared to the limits in Figure~\ref{fig:Cggbounds}. If the FCC-ee will operate at different values of the center-of-mass energy, it is in principle possible to measure the two coefficients $C^\text{eff}_{\gamma Z}$ and $C^\text{eff}_{\gamma\gamma}$ independently, as pointed out in \cite{Bauer:2017ris}. Also, for the proposed $Z$-pole run of the FCC-ee, the process $e^+e^-\to \gamma a\to 3 \gamma$ would correspond to on-shell decay of the $Z$ boson to an ALP, $Z\to \gamma a$, which will be discussed below.

\begin{figure}[t]
\begin{center}
\includegraphics[width=\textwidth]{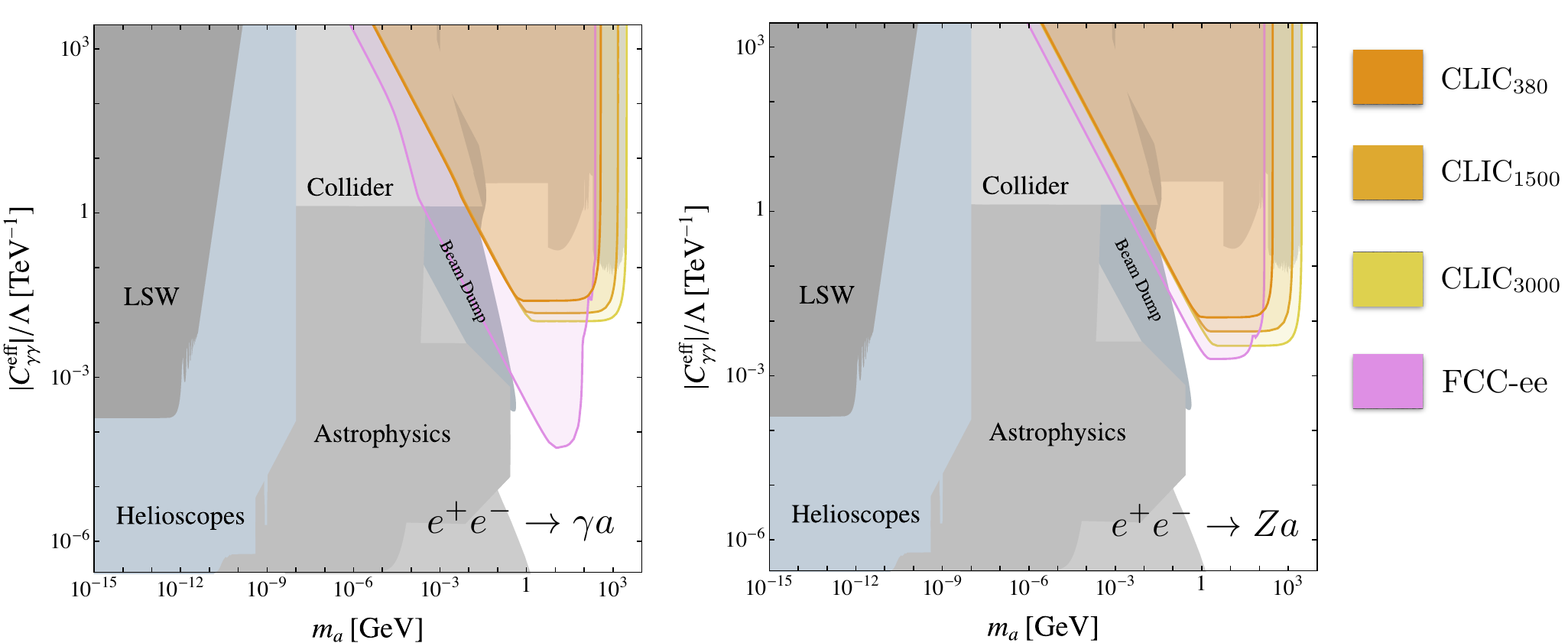}
\end{center}
\vspace{-3mm}
\caption{\label{fig:cggboundsee} Projected sensitivity regions for searches for $e^+e^-\to \gamma a \to 3\gamma$ (left) and $e^+e^-\to Z a \to Z_\text{vis}\gamma
\gamma$ (right) at future $e^+e^-$ colliders for $\text{Br}(a\to\gamma
\gamma)=1$. The constraints from Figure~\ref{fig:Cggbounds} are shown in the background. The sensitivity regions are based on 4 expected signal events.
}
\end{figure}

We show the projections for the various versions of the CLIC collider and the FCC-ee in Figure~\ref{fig:cggboundsee}, assuming $C_{WW}=0$ which implies $C_{\gamma Z}=-s_w^2C_{\gamma\gamma}$.\footnote{Note that the assumption Br$(a\to \gamma\gamma)=1$ is not justified for $m_a> m_Z$, for which the decay channel $a\to Z \gamma$ opens up. Even though this corresponds to a different final state, we expect similarly effective cuts for $a \to Z\gamma$ and do not treat this final state differently in Fig. \ref{fig:cggboundsee}.} The parameter space corresponds to at least 4 expected signal events with the ALP decaying before it has reached the electromagnetic calorimeter (ECAL) which is assumed to be within a radius of $\sim 1.5\,$m of the beam axis. We consider only visible decays of the $Z$ boson with $\text{Br}(Z\to \text{visible})=0.80$. We also impose the constraint $|C_{\gamma Z}^{\rm eff}| < 1.48\, \Lambda/\text{TeV} $ from the LEP measurement of the 
total width of the $Z$ boson.

%
%

%
\begin{figure}[t]
\begin{center}
\includegraphics[width=\textwidth]{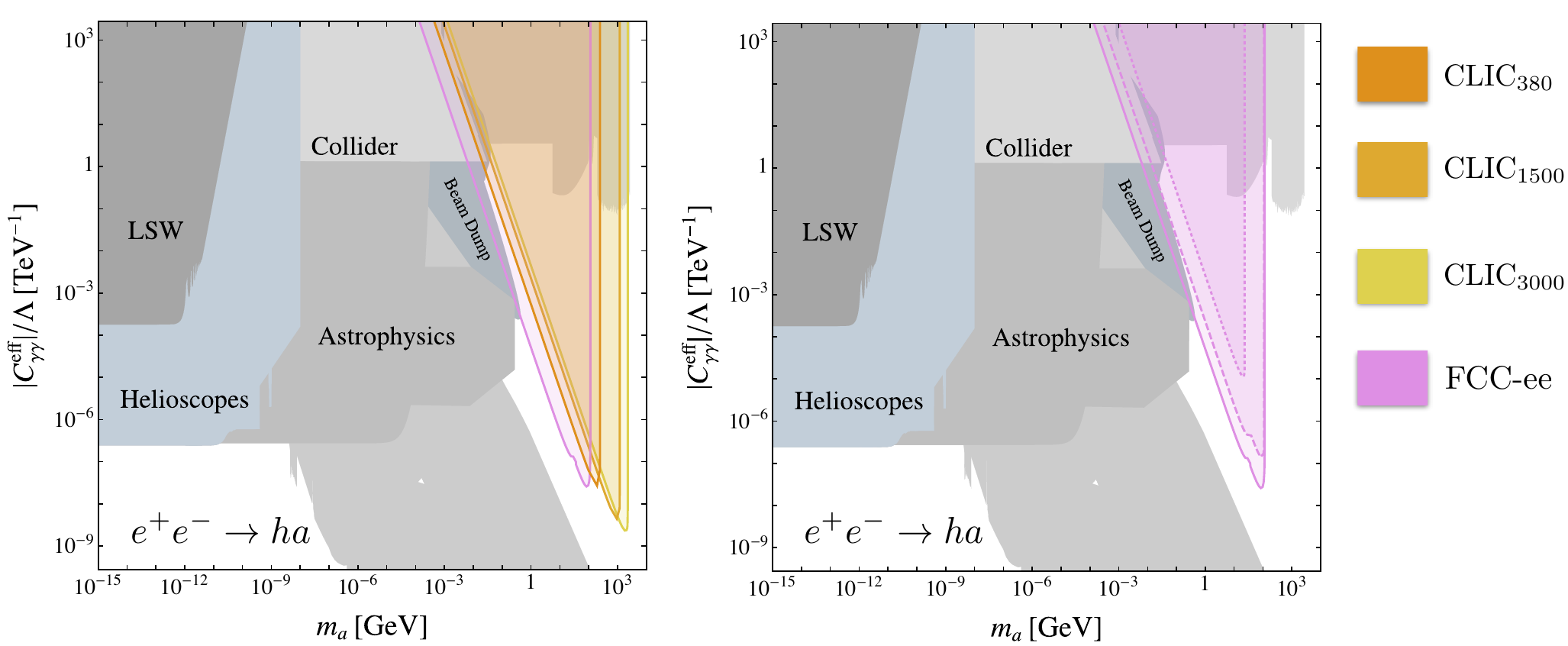}
\includegraphics[width=\textwidth]{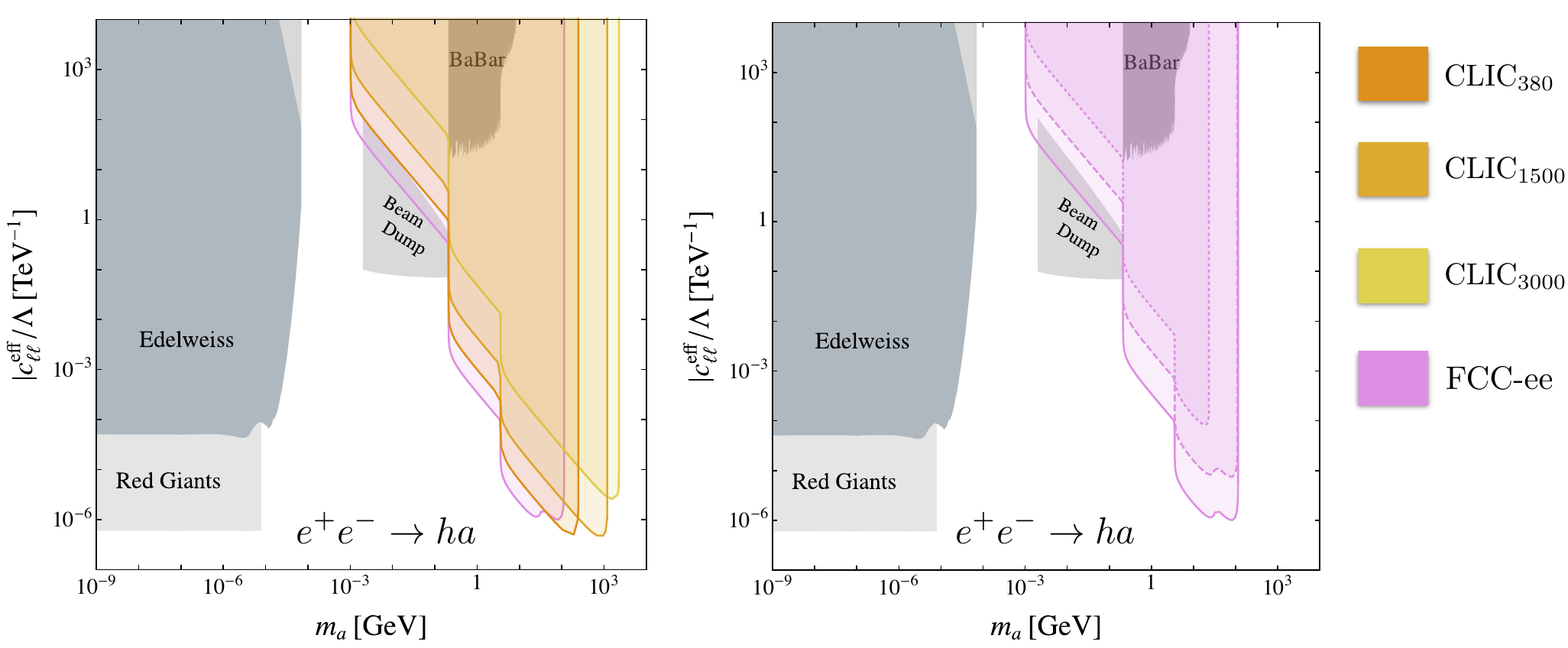}
\end{center}
\vspace{-3mm}
\caption{\label{fig:eehabounds} Left: Projected sensitivity regions for searches for $e^+e^-\to ha \to b\bar b \gamma\gamma$ (upper panels) and $e^+e^-\to ha \to b\bar b \ell^+\ell^-$ (lower panels) for future $e^+e^-$ colliders, assuming that $|C_{Zh}^\text{eff}|=0.72\,\Lambda/\text{TeV}$ and $\text{Br}(a\to\gamma
\gamma)=1$ (upper panels) and $\text{Br}(a\to\ell^+\ell^-)=1$ (lower panels). Right: Sensitivity regions for the example of the FCC-ee with $|C_{Zh}^\text{eff}|=0.72\,\Lambda/\text{TeV}$ (solid contour), $|C_{Zh}^\text{eff}|=0.1\,\Lambda/\text{TeV}$  
(dashed contour), and $|C_{Zh}^\text{eff}|=0.015\,\Lambda/\text{TeV}$ (dotted contour), which corresponds to Br$(h \to Za)=34\%, 1\% $ and Br$(h \to Za)=0.02\% $, respectively.  The constraints from Figure~\ref{fig:Cggbounds} are shown in the background. The sensitivity regions are based on 4 expected signal events.}
\end{figure}
%

The contours for the FCC-ee in Figure~\ref{fig:cggboundsee} combine the luminosities for the run at the $Z$-pole (in case of $e^+e^-\to  \gamma a$), at $\sqrt{s}=2m_W$ and at $\sqrt{s}=250\,$GeV, whereas for CLIC we show separate limits for three different versions of this collider. Note that 
the large luminosity of the FCC-ee run at the $Z$ pole leads to a significantly larger sensitivity in the $e^+e^-\to \gamma a$ channel compared to the $e^+e^-\to 
Z a $ projection. Further, CLIC$_\text{1500}$ and CLIC$_{3000}$ allow to probe  considerably higher ALP masses compared to both CLIC$_{380}$ and the FCC-ee. In this and the following figures, the relevant ALP branching ratio into the observed final state is set to a 100\%. As we have shown in \cite{Bauer:2017ris}, the left boundary of the sensitivity region is largely independent of this assumption. For branching ratios smaller than $\text{Br}(a\to \gamma\gamma)=1$, the reach in $C_{\gamma\gamma}^\text{eff}$ however is reduced by a factor $\big[\text{Br}(a\to \gamma\gamma)\big]^{1/2}$. This follows from the cross sections \eqref{sigmahighE1} and \eqref{sigmahighE2}, which imply the scaling $\sigma (e^+e^-\to \gamma a\to 3\gamma) \sim |C_{\gamma\gamma}^\text{eff}|^2\,\text{Br}(a\to \gamma\gamma)$ and  $\sigma (e^+e^-\to Z a\to Z \gamma\gamma) \sim |C_{\gamma\gamma}^\text{eff}|^2\,\text{Br}(a\to \gamma\gamma)$, respectively.\footnote{Here we have again used that $C_{\gamma Z}=-s_w^2C_{\gamma\gamma}$.}

ALPs can also be produced in association with a Higgs boson. The rate for the process $e^+e^- \to h a$ depends on the Wilson coefficient $C_{Zh}^\text{eff}$ in 
\eqref{LeffD>5}. The constraint $\Gamma(h\to \text{BSM})< 2.1\,$MeV on the partial Higgs decay width into non-SM final states implies the upper bound $|C^\text{eff}_{Zh}|< 0.72\,\Lambda/\text{TeV}$ \cite{Khachatryan:2016vau}. Assuming that the Higgs boson is reconstructed in the $b\bar b$ final states with $\text{Br}(h \to b\bar b)= 0.58$, we derive the sensitivity to $C_{\gamma\gamma}^{\rm eff}$ and $m_a$ displayed in the upper left panel of Figure~\ref{fig:eehabounds}. 
In the upper right panel of Figure~\ref{fig:eehabounds} we show how these projected sensitivity regions vary for different values of $C^\text{eff}_{Zh}$. The expected sensitivity remains the same down to a critical value of the branching ratio $\text{Br}(a\to \gamma\gamma)<1$.  Below this critical value less than 4 events are produced and the discovery reach is lost. For the FCC-ee, these critical values are $\text{Br}(a\to\gamma\gamma)=2\times 10^{-4}$ for $C_{Zh}^\text{eff}=0.72 \, \Lambda/\text{TeV}$, $\text{Br}(a\to\gamma\gamma)=10^{-2}$ for $C_{Zh}^\text{eff}=0.1 \, \Lambda/\text{TeV}$ and $\text{Br}(a\to\gamma\gamma)=0.4$ for $C_{Zh}^\text{eff}=0.015 \, \Lambda/\text{TeV}$. For the case of leptonic ALP decays these values do not change, and they are only slightly different in the case of CLIC. In that 
case, searches for other final states can become more promising. This includes searches for invisibly decaying (or stable) ALPs \cite{Dolan:2017osp}.
Lepton colliders are particularly powerful in constraining ALP-lepton couplings. In order to avoid large lepton-flavor changing ALP couplings, we choose a benchmark with  ALP couplings to leptons,  
\begin{align}
c_{\ell\ell}\equiv c_{ee}=c_{\mu\mu}=c_{\tau\tau}\,.
\end{align}
The lower panels of Figure~\ref{fig:eehabounds} show the regions of sensitivity for ALP searches in the process $e^+e^-\to ha\to b\bar b\, \ell^+\ell^-$. The jumps in the sensitivity region appear at the thresholds for the production of muon and tau pairs.  The ALP decays predominantly into the heaviest lepton that is kinematically accessible.

%
\begin{figure}[t]
\begin{center}
\includegraphics[width=.9\textwidth]{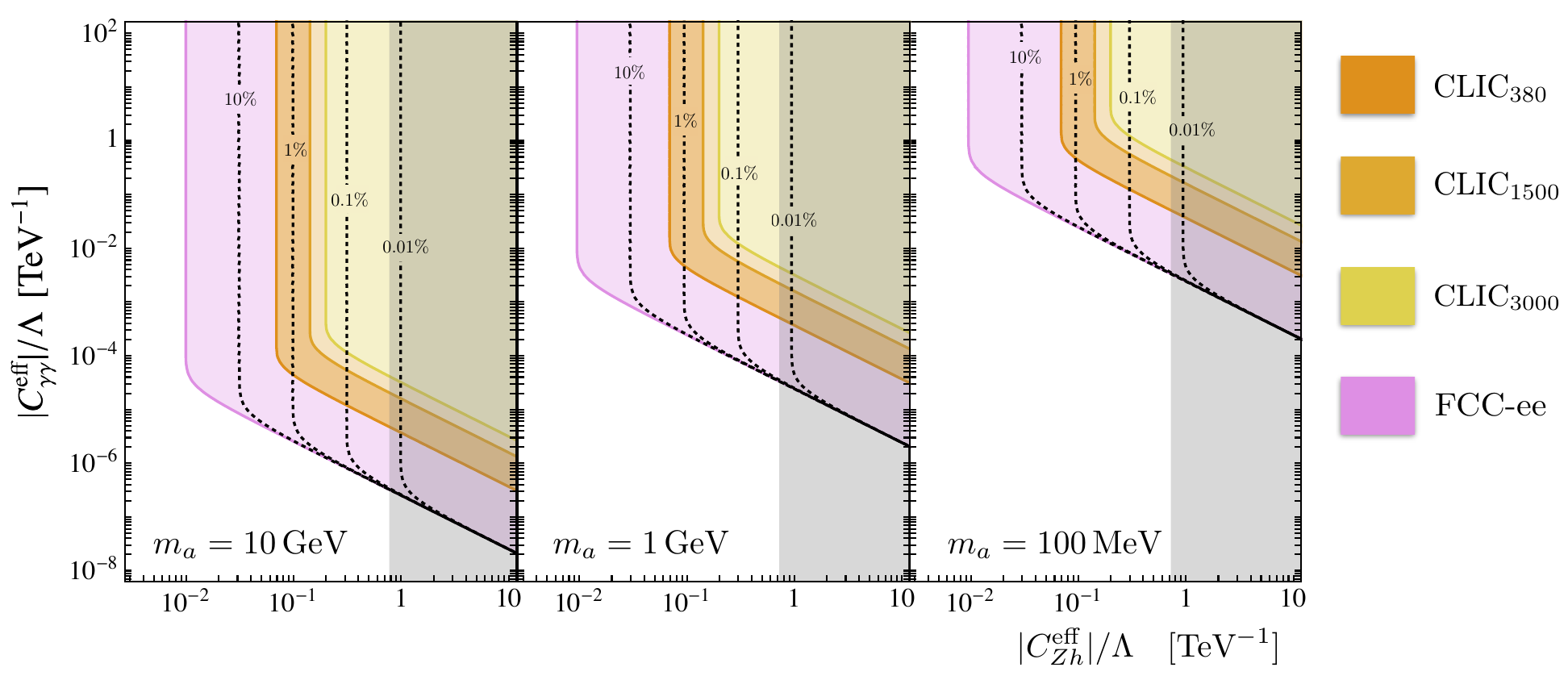} \\
\includegraphics[width=.9\textwidth]{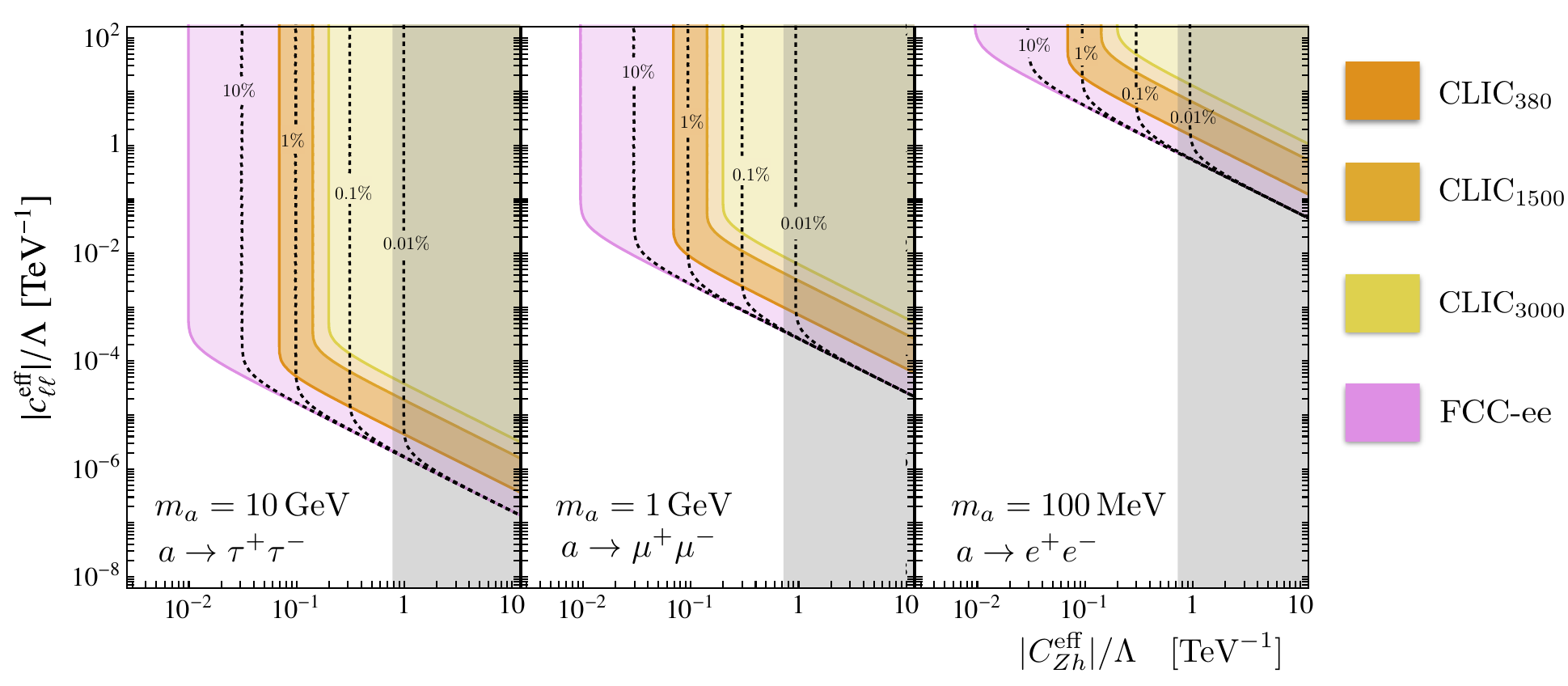}
\end{center}
\vspace{-.8cm}
\caption{\label{fig:higgsdecaysL_couplingParam} Parameter regions which can be probed for $e^+e^-\to ha \to b\bar b \gamma\gamma$ (upper panels) and $e^+e^-\to ha \to b\bar b \ell^+\ell^-$ (lower panels) at future $e^+e^-$ colliders. The grey shaded area is excluded by LHC Higgs measurements. The parameter space to the right of the dotted contours corresponds to the sensitivity reach of the FCC-ee with the indicated ALP branching ratios. The sensitivity regions are based on 4 expected signal events.}
\end{figure}
%

The graphical representation in Figure~\ref{fig:eehabounds} is suboptimal, because it highlights the dependence on one ALP coupling ($|C_{\gamma\gamma}^\text{eff}|$ or $|c_{\ell\ell}^\text{eff}|$), while the dependence on the other coupling ($C_{Zh}^\text{eff}$) is only reflected by the different contours. In Figure~\ref{fig:higgsdecaysL_couplingParam} we show an alternative representation of the results in the plane of the two relevant ALP couplings, but for fixed values of the ALP mass. The sensitivity reach of the FCC-ee and the three versions of the CLIC collider for an ALP branching ratio of $\text{Br}(a\to \gamma\gamma)=1$ (upper panels) and $\text{Br}(a\to \ell^+\ell^-)=1$ (lower panels) is bounded by the coloured contours. 
With decreasing ALP mass, the lifetime of the ALP increases and the sensitivity reach in $C_{\gamma\gamma}^\text{eff}$ and $c_{\ell\ell}^\text{eff}$ is reduced. The fact that the sensitivity region for CLIC is maximal for the lowest center-of-mass energy is a consequence of the $1/s$ behaviour of the $e^+e^-\to ha$ cross section in \eqref{sigmahighE3}.

For the example of the FCC-ee, we also indicate the dependence of the sensitivity regions on the $a\to \gamma\gamma$ or $a\to \ell^+\ell^-$ branching ratios, which in Figure~\ref{fig:eehabounds} were assumed to be maximal. The parameter space to the right of the dotted contours corresponds to the sensitivity reach of the FCC-ee with the indicated ALP branching ratios. Smaller branching ratios reduce the sensitivity to $C_{Zh}^\text{eff}$, because the total number of signal events decreases. However, the values of $C_{\gamma\gamma}^\text{eff}$ and $c_{\ell\ell}^\text{eff}$ for which sensitivity is lost are almost independent of the ALP branching ratio, as long as this branching ratio exceeds a critical value. Consider, for example the process $e^+e^-\to ha\to b\bar b\gamma\gamma$ for $m_a=10\,$GeV (upper left panel of Figure~\ref{fig:higgsdecaysL_couplingParam}). If  $C_{Zh}^\text{eff}/\Lambda=0.1\,\text{TeV}^{-1}$, the sensitivity reach in $C_{\gamma\gamma}^\text{eff}/\Lambda$ extends down to $\approx 10^{-5}\,\text{TeV}^{-1}$ irrespective of $\text{Br}(a\to\gamma\gamma)$, as long as this branching ratio exceeds 1\%.  The reason for this behaviour is that the total width of the ALP increases for smaller ALP branching ratios and therefore the lifetime decreases. Smaller ALP lifetimes lead to more ALP decays in the detector volume, canceling the effect of the reduced branching ratio near the lower boundary of the sensitivity region \cite{Bauer:2017ris}. 
In order to not clutter the plots we do not show the corresponding contours for CLIC. 

From now on, whenever ALP production and decay are governed by unrelated Wilson coefficients, we will use the graphical representation in Figure~\ref{fig:higgsdecaysL_couplingParam}.

A particularly interesting benchmark scenario is the model in which at tree-level the ALP only couples to charged leptons. In this case the production and decay are governed by the same parameter $c_{\ell\ell}$. The
ALP decays are dominated by $\text{Br}(a\to e^+e^-)\approx 1$ for $m_a< 2m_\mu$, $\text{Br}(a\to \mu^+\mu^-)\approx 1$ for $2m_\mu<m_a< 2m_\tau$, and $\text{Br}(a\to \tau^+\tau^-)\approx 1$ for $m_a> 2m_\tau$. Interestingly, the most relevant production mode at $e^+e^-$ colliders is still the associated production with photons and $Z$ bosons, which proceeds through the loop-induced Wilson coefficients \cite{Bauer:2017ris}
\begin{align}
C_{\gamma \gamma}^\text{eff} &= \frac{1}{16\pi^2}\, c_{\ell\ell}\, \sum_{\ell=e, \mu, \tau}B_1(\tau_\ell)\,,\\
C_{\gamma Z}^\text{eff}&=\frac{1}{16\pi^2}\Big(s_w^2-\frac{1}{4}\Big) \,c_{\ell\ell}\,\sum_{\ell=e, \mu, \tau} B_3(\tau_\ell, \tau_{\ell/Z})\approx\Big(s_w^2-\frac{1}{4}\Big)\,C^\text{eff}_{\gamma\gamma}\,,
\end{align}
with $\tau_\ell=4m_\ell^2/m_a^2$, and $\tau_{\ell/Z}=4m_\ell^2/m_Z^2$. In the last step in the second equation we have neglected terms of order $m_\ell^2/m_Z^2$. Because of the anomaly equation, $B_1(\tau_\ell)\approx 1$ for $m_a>m_\ell$ and $B_1(\tau_\ell)\approx -\frac{m_a^2}{12m_\ell^2}$ for $m_\ell \gg m_a$ and the relative size of the resonant production cross section and the associated ALP+$\gamma$ production cross section is given by 
\begin{align}
\frac{\sigma(e^+e^-\to \gamma a)}{\sigma(e^+e^-\to a )}&= \frac{\alpha \,\alpha(s)^2}{12\pi^2}\,N_{\ell}^2\,\frac{s^2}{\Gamma_a m_a m_e^2 }\bigg(1-\frac{m_a^2}{s}\bigg)^5\notag\\&\approx 1.3\times 10^{11}\,\bigg[\frac{N_\ell}{3}\bigg]^2\bigg[\frac{s}{\text{TeV}}\bigg]^2 \bigg[\frac{\text{GeV}}{m_a}\bigg] \bigg[\frac{\text{keV}}{\Gamma_a}\bigg]\,,
\end{align}
where $N_\ell$ denotes the number of charged leptons lighter than the ALP, and $\Gamma_a\approx $ keV is a typical width for $a\to \tau^+\tau^-$, assuming $|c_{\ell\ell}|/\Lambda \approx 1/\text{TeV}$. For $N_\ell<3$, the total width is reduced by $m_\mu^2/m_\tau^2$, and the associated ALP$+\gamma$ production is even more dominant.   
The ratio of the partial decay widths on the other hand is given by 
\begin{align}
\frac{\Gamma(a\to \ell\ell)}{\Gamma(a\to \gamma\gamma)}\approx \frac{8 \pi^2 m_\ell^2}{\alpha^2 m_a^2 N_\ell^2}  \approx 4.1\times 10^4\,\bigg[\frac{3}{N_\ell}\bigg]^2\frac{4m_\ell^2}{m_a^2} \,,
\end{align}
with $m_\ell$ the mass of the heaviest lepton in which the ALP can decay. For ALP masses below 720\,GeV (2300\,GeV) this ratio is larger than 1 (0.1), justifying the assumption of $\text{Br}(a\to \ell^+\ell^-)=1$ for almost all of the relevant parameter space. 

%
\begin{figure}[t]
\begin{center}
\includegraphics[width=\textwidth]{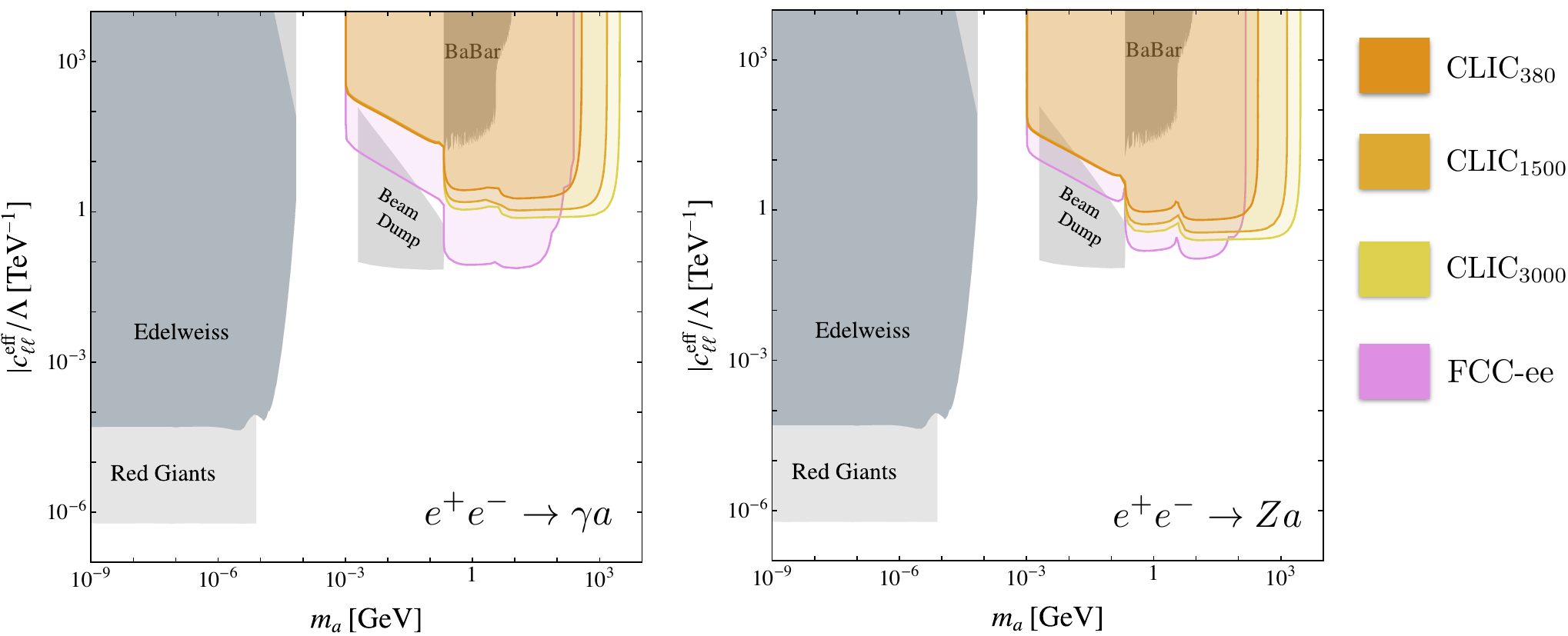}
\end{center}
\vspace{-3mm}
\caption{\label{fig:eebounds} Projected exclusion contours for searches for $e^+e^-\to \gamma a \to \gamma \ell^+\ell^-$ (left) and $e^+e^-\to Z a \to Z_\text{vis}\ell^+
\ell^-$ (right) for future $e^+e^-$ colliders, and $\text{Br}(a\to\ell^+
\ell^-)=1$. The constraints from Figure~\ref{fig:Cllbounds} are in the background. The sensitivity regions are based on 4 expected signal events.
}
\end{figure}
%

We show projections for future $e^+e^-$ colliders for flavor universal ALP-lepton couplings in Figure~\ref{fig:eebounds}. An increase in sensitivity occurs at the di-muon and di-tau thresholds. Note that while the advantage of a high-luminosity run on the $Z$-pole of the FCC-ee accounts for an increase in sensitivity on $C_{\gamma\gamma}^\text{eff}$ of up to $\sim 2.5$ orders of magnitude in Figure~\ref{fig:cggboundsee}, for purely leptonic ALP couplings the $Z$-pole run only increases the sensitivity by about one order of magnitude in $e^+e^-\to \gamma a$, because the loop-induced Wilson coefficient $C_{\gamma Z}^{\rm eff}$ is suppressed by the accidentally small vector coupling of the $Z$ boson to charged leptons. CLIC can again constrain higher ALP masses.

\subsubsection*{\boldmath ALP production in exotic decays of on-shell Higgs bosons}

%
\begin{figure}[t]
\begin{center}
\includegraphics[width=.9\textwidth]{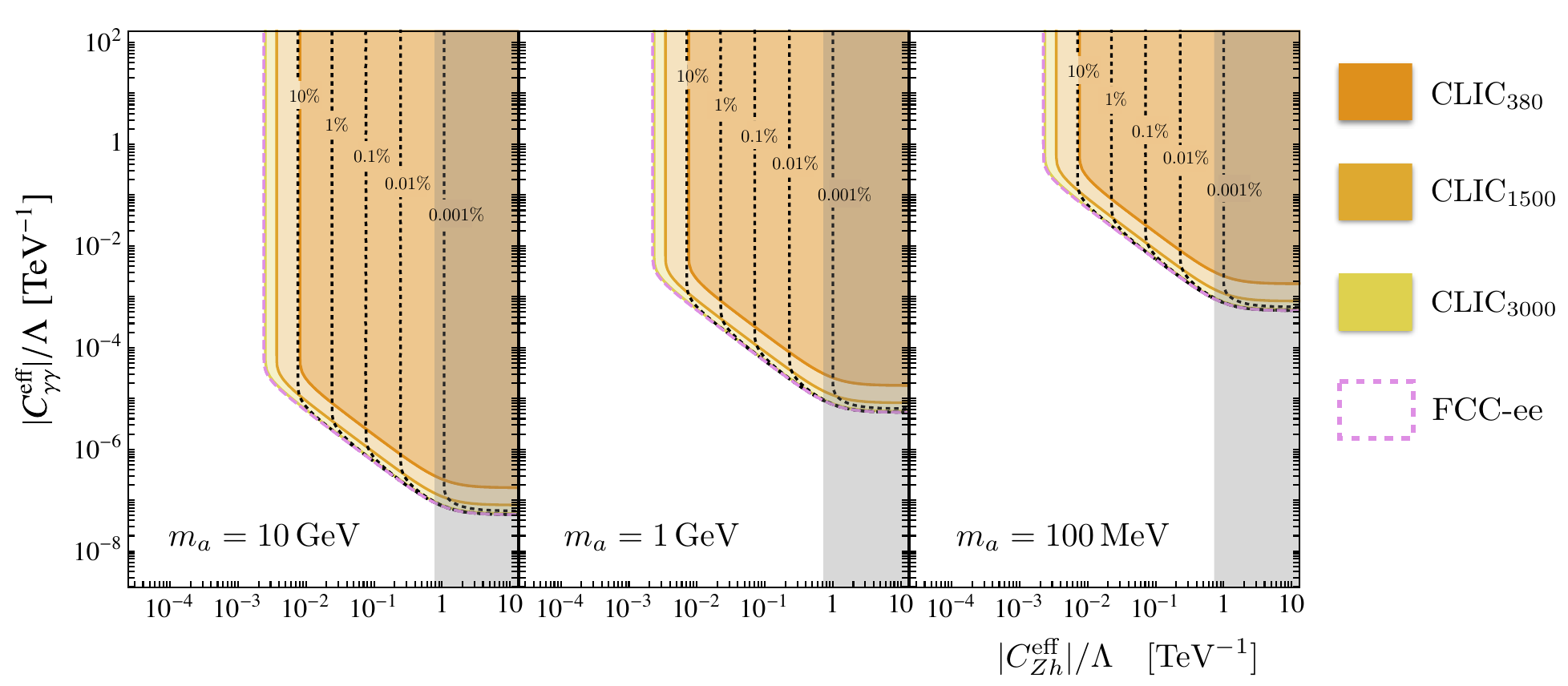} \\
\includegraphics[width=.9\textwidth]{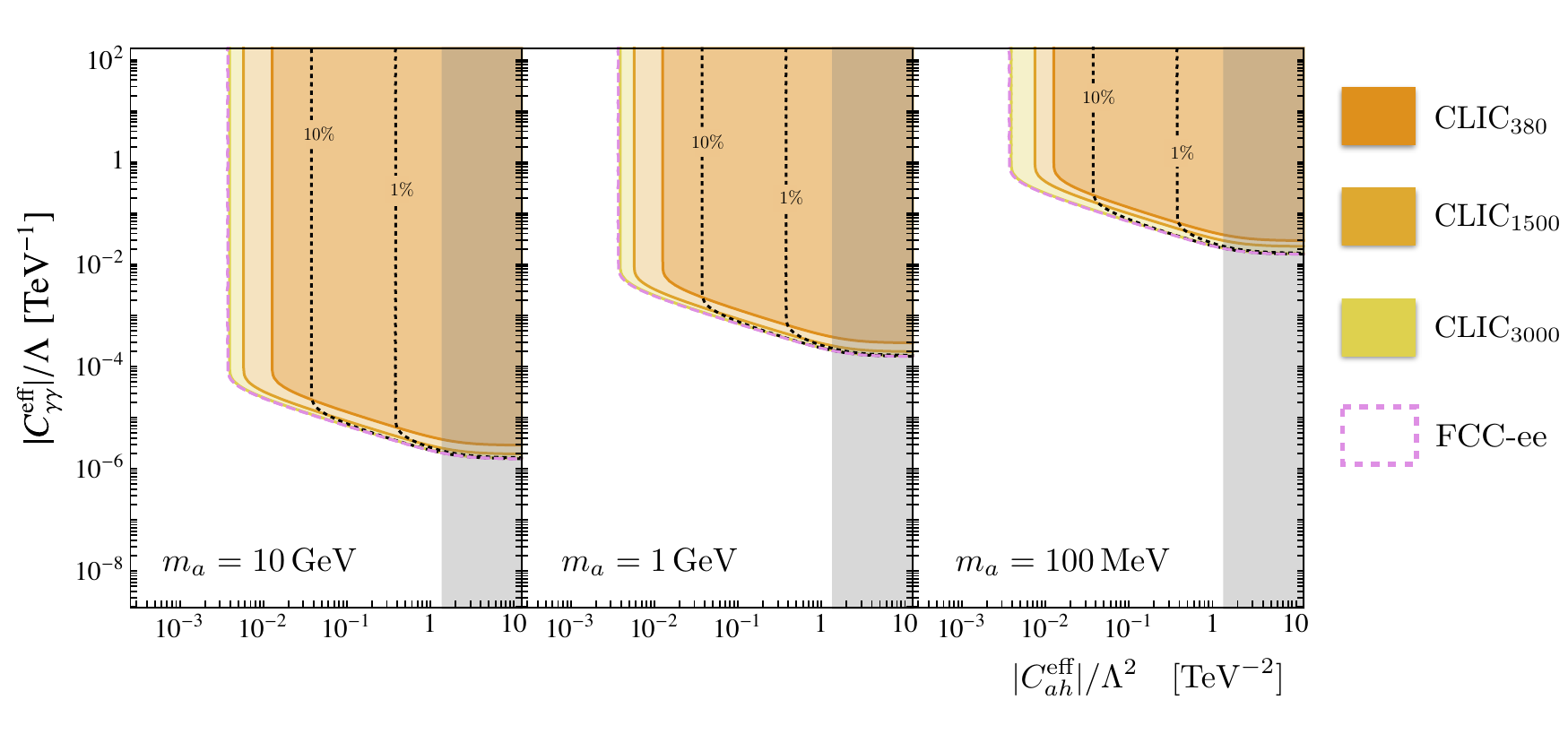}
\end{center}
\vspace{-.8cm}
\caption{\label{fig:higgsdecaysL} Parameter regions which can be probed in the decay $h \to Z a$ with $a \to \gamma \gamma$ (upper row) and $h \to a a$ with $a \to \gamma \gamma$ (lower row) at future $e^+e^-$ colliders. The grey shaded area is excluded by LHC Higgs measurements. The dotted contours correspond to the sensitivity region of the FCC-ee for ALP branching ratios smaller than 1. The sensitivity regions are based on 4 expected signal events.}
\end{figure}
%

%
\begin{figure}[t]
\begin{center}
\includegraphics[width=.9\textwidth]{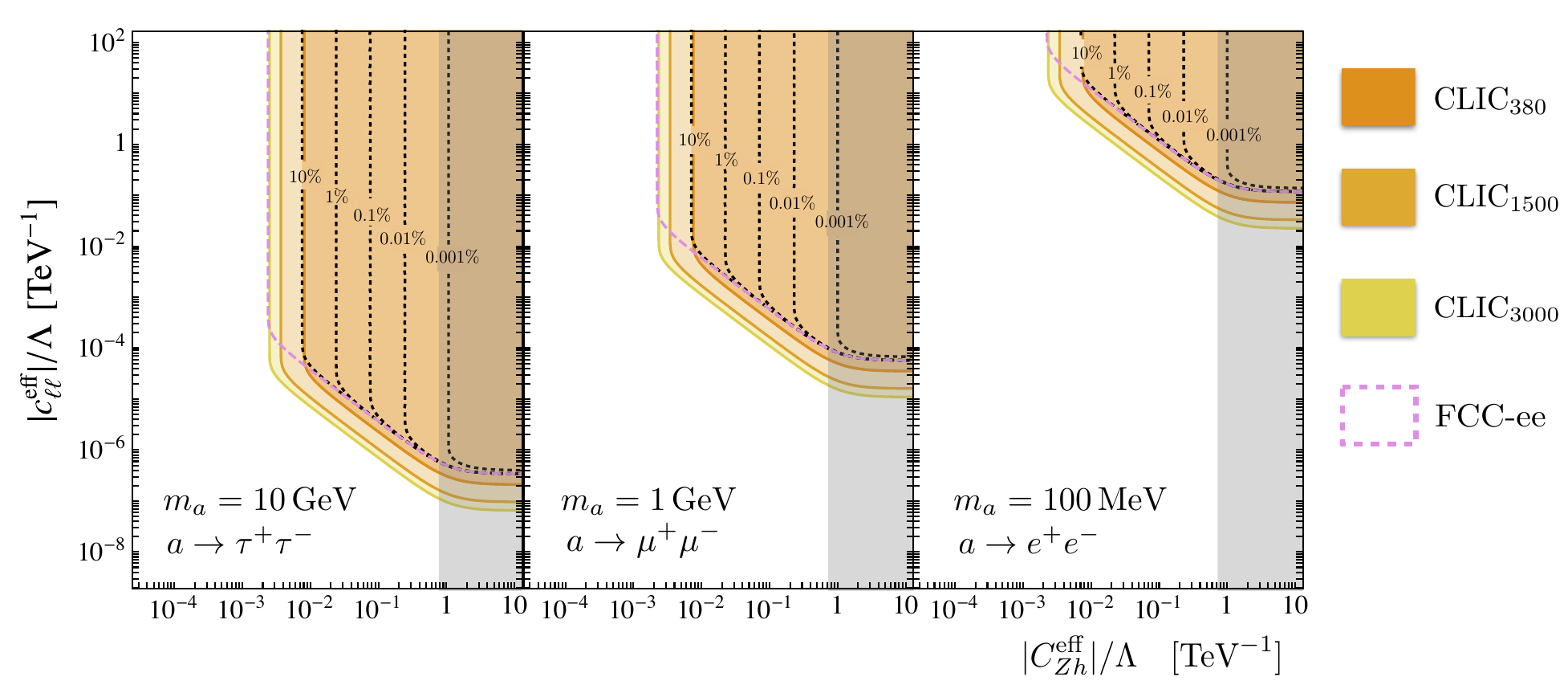} \\
\hspace{-.2cm}\includegraphics[width=.9\textwidth]{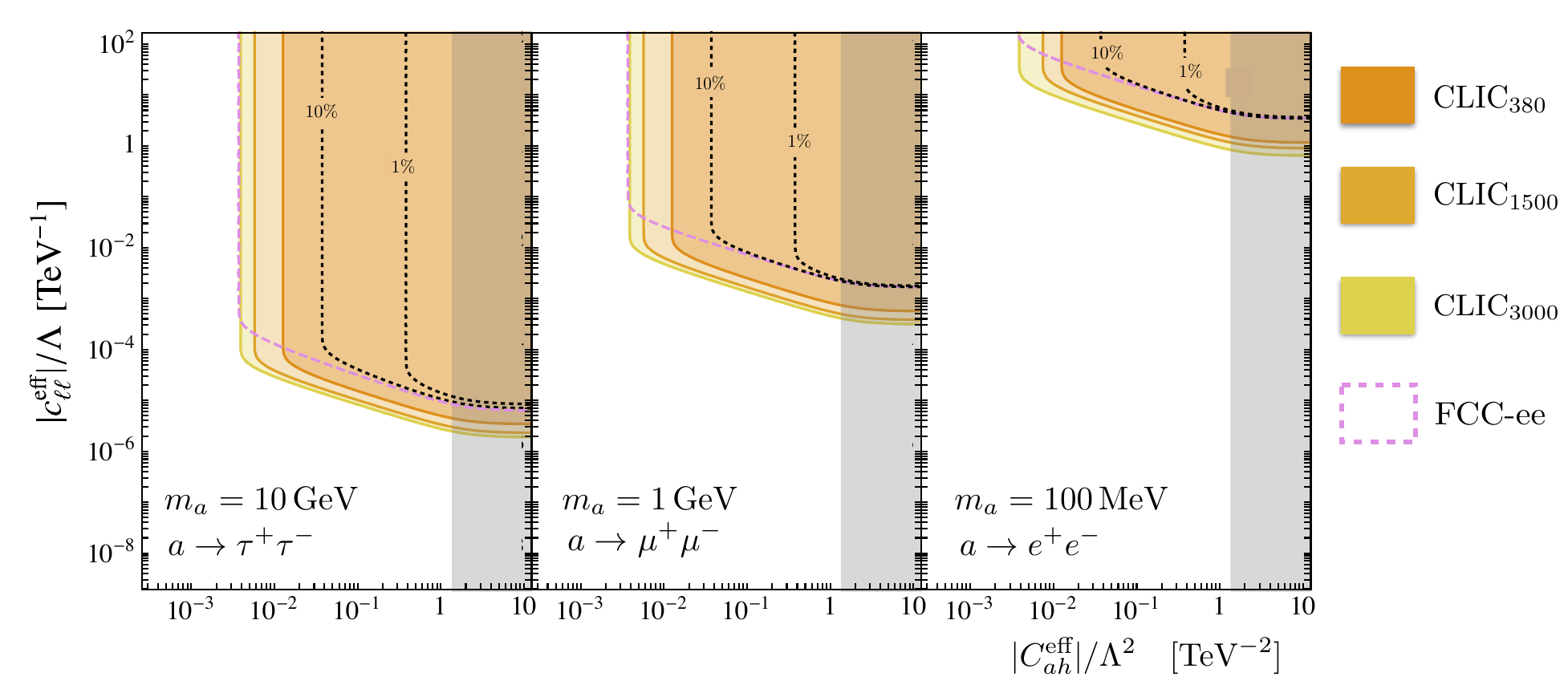}
\end{center}
\vspace{-.3cm}
\caption{\label{fig:higgsdecaysLlep} Parameter regions which can be probed in the decay $h \to Z a$ with $a \to \ell^- \ell^+$ (upper row) and $h \to a a$ with $a \to \ell^+ \ell^-$ (lower row) at future $e^+e^-$ colliders. The grey shaded area is excluded by LHC Higgs measurements. The dotted contours correspond to the sensitivity region of the FCC-ee for ALP branching ratios smaller than 1. The sensitivity regions are based on 4 expected signal events.}
\end{figure}
%

%
\begin{figure}[t]
\begin{center}
\includegraphics[width=0.5\textwidth]{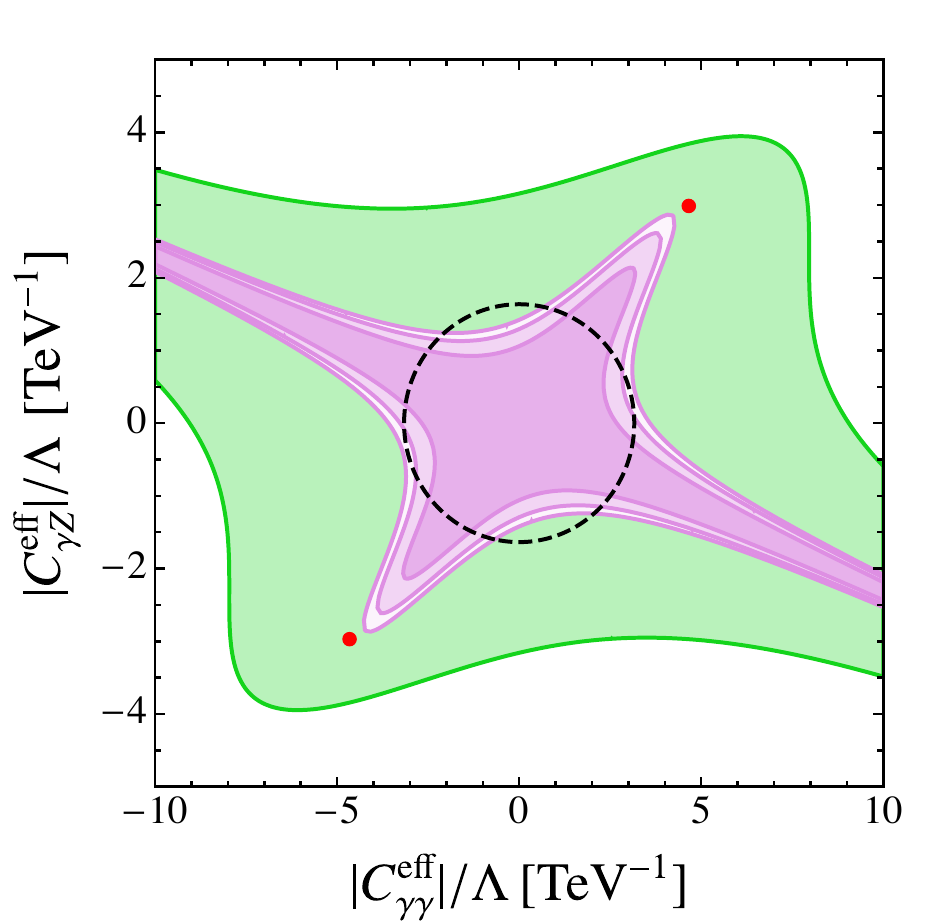}
\end{center}
\vspace{-3mm}
\caption{\label{fig:EWPT} Allowed regions in the parameters space of the Wilson coefficients $C_{\gamma\gamma}^{\rm eff}-C_{\gamma Z}^{\rm eff}$ obtained from projections 
for the two-parameter global electroweak fit at 68\% CL, 95\% CL  and 99\% CL at FCC-ee (violet) and at 95\% CL for the LHC at $\sqrt{s}=14\,$TeV (green). For the parameter space within the dashed black contour, 
the FCC-ee measurement of $\alpha(m_Z)$ is within its projected errors at 95\% CL \cite{Janot:2015gjr}. The red dots represent the best fit points based on the current electroweak fit.}
\end{figure}
%

Beyond searches for ALPs produced in association with a photon, a $Z$ boson or a Higgs boson, ALPs can also be searched for in exotic Higgs decays. The Higgs 
production cross section at lepton colliders is typically at least one order of magnitude smaller compared to the LHC. This implies that lepton colliders are most powerful 
for light ALPs with dominant decay channels for which backgrounds at hadron colliders are large.  
In Figure~\ref{fig:higgsdecaysL}, we show the reach of the different stages of CLIC and the FCC-ee for ALPs produced in $e^+e^- \to h+X \to a Z +X \to \gamma
\gamma Z_\text{vis}+X$ and $e^+e^- \to h + X\to a a + X\to 4\gamma+X$ for three different ALP masses $m_a= 100\,$MeV, $1\,$GeV and $10\,$GeV. We do not distinguish between vector-boson fusion or associated Higgs production and 
demand four signal events. In order to reconstruct the Higgs, we further demand the $Z$ boson to originate from the Higgs decay as well as all $Z$s to decay into
visible final states with $\text{Br}(Z\to \text{visible})=0.8$ and $\text{Br}(a\to \gamma\gamma)=1$. This condition can be relaxed if the electrons in $ZZ$-fusion or 
the additional $Z$ in associated Higgs production are detected. Since the reach in searches for exotic Higgs decays is directly proportional to the number of 
Higgses produced, high-luminosity machines lead to the best sensitivity. In Figure~\ref{fig:higgsdecaysL} we further show the reach of the FCC-ee for different values of $\text{Br}(a\to \gamma\gamma)=10^{-5}-10^{-1}$ given by the respective dotted lines. For leptonic ALP decays, the analogous plots are shown in Figure~\ref{fig:higgsdecaysLlep}, where, in contrast to Figure~\ref{fig:eebounds}, no connection between $C_{ah}^\text{eff}$, $C_{Zh}^\text{eff}$ and $c_{\ell\ell}^{\rm eff}$ has been assumed. 
CLIC has a larger reach than the FCC-ee for leptonic ALP decays due to the larger detector volume, $L_\text{det}=0.6\,$m at CLIC, compared to $L_\text{det}=0.02\,$m at the FCC-ee.
 Since $C_{ah}^\text{eff}$ and $C_{Zh}^\text{eff}$ are not controlled by the anomaly equation, the one-loop contribution from a tree-level $c_{\ell\ell}^{\rm eff}$ coupling is proportional to $m_\ell^2/v^2$ \cite{Bauer:2017ris}. The grey regions in Figures~\ref{fig:higgsdecaysL} and \ref{fig:higgsdecaysLlep} correspond to $|C_{Zh}^\text{eff}| >0.72  \Lambda/\text{TeV}$  and $|C_{ah}^\text{eff} | >1.34\,\Lambda^2/\text{TeV}^2$ excluded by the current upper limit on $\text{Br}(h \to \text{BSM})< 0.34$ (at 95\% CL) \cite{Khachatryan:2016vau}.

\subsubsection*{\boldmath Electroweak precision constraints on ALP couplings}

Besides direct measurements, lepton colliders will be able to measure electroweak observables with unprecedented precision, which allows us to set bounds on the ALP contributions to these observables \cite{Bauer:2017ris}. The measurement of the oblique parameters will improve current constraints by roughly one order of magnitude \cite{deBlas:2016ojx}, while the running of the 
electromagnetic coupling constant, $\alpha(m_Z)$, can be determined with an uncertainty of about $10^{-5}$ \cite{Janot:2015gjr}.
In Figure~\ref{fig:EWPT}, we show the projected electroweak fit for the FCC-ee, where we assume the central values to correspond to the SM prediction, in the $C_{\gamma\gamma}^{\rm eff}-C_{\gamma Z}^{\rm eff}$ plane
at 68\% , 95\%  and 99\% CL (violet), together with the expected sensitivity of the LHC at $\sqrt{s}=14\,$TeV (green). Superimposed is the expected 95\% CL bound derived from the measurement of $\alpha(m_Z)$ (black 
dashed contour), assuming that the theoretical error on this quantity will have decreased below the experimental uncertainty by the time the measurement can be 
performed. In deriving these projections we have set the ALP mass to zero. 
By combining the future measurements of $\alpha(m_Z)$ and of electroweak precision pseudo-observables one will be able to constrain $|C_{\gamma\gamma}^{\rm eff}|/\Lambda \lesssim 2.5\,$TeV$^{-1}$ and $|C_{\gamma Z}^{\rm eff}|/\Lambda \lesssim 1.5\,$TeV$^{-1}$ (at 95\% CL).
The current global fit has a slight tension with the SM prediction and the best fit point is at $(S,T)= (0.096, 0.111)$. If this effect is solely due to the ALP couplings $C_{\gamma\gamma}^{\rm eff}$ and $C_{\gamma Z}^{\rm eff}$, the corresponding best fit points are indicated by the red dots in Figure~\ref{fig:EWPT}. Such sizeable coefficients are however strongly constrained by LHC searches for $pp\to \gamma a$ and $pp\to  \gamma Z$.

\subsection{ALP searches at future hadron colliders}\label{sec:hadcol}

Future hadron colliders can significantly surpass the reach of the LHC in searches for ALPs.
In particular, searches for ALPs produced in exotic Higgs and $Z$ decays profit from the higher center-of-mass energies and luminosities of the proposed high-energy LHC (HE-LHC), planned to replace the LHC in the LEP tunnel with $\sqrt{s}=27 \,$TeV, and the ambitious plans for a new generation of hadron colliders with $\sqrt{s}=100\,$TeV at CERN (FCC-hh) and in China (SPPC). At hadron colliders, ALP production in association with electroweak bosons suffers from large backgrounds. Previous studies of these processes have therefore focussed on invisibly decaying (or stable) ALPs, taking advantage of the missing-energy signature \cite{Mimasu:2014nea,Brivio:2017ije}. In contrast, here we focus our attention on resonant ALP production in gluon-fusion and photon-fusion, as well as on ALPs produced in the decays of $Z$ and Higgs bosons. 

%
\begin{figure}[t]
\begin{center}
\includegraphics[width=0.9\textwidth]{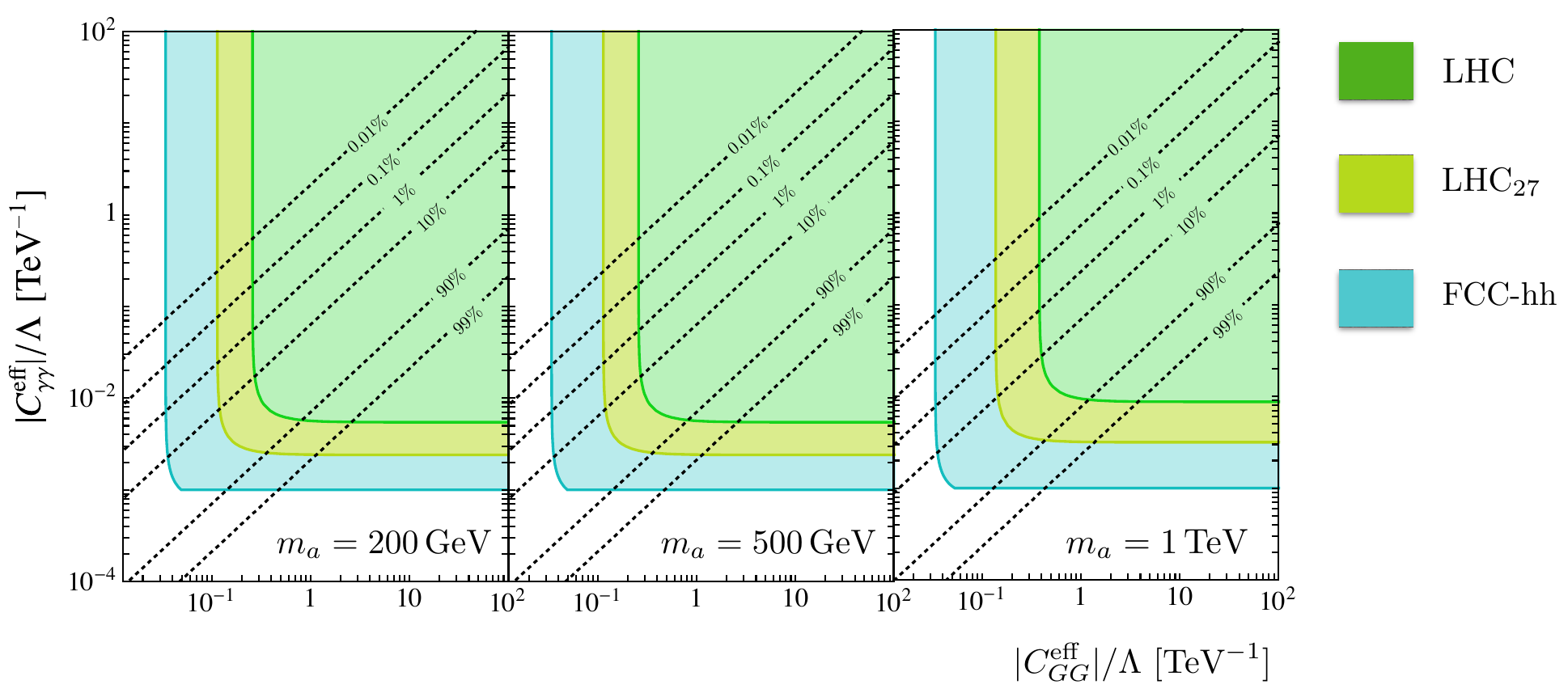}
\end{center}
\vspace{-3mm}
\caption{\label{fig:gluonfusion} Projected reach in searches for $pp \to a \to \gamma\gamma $ with the LHC 
(green), HE-LHC (light green) and a $100\,$TeV collider (blue). Contours of constant branching ratios $\text{Br}(a\to\gamma\gamma)$ are shown as dotted lines. The sensitivity regions are based on 100 expected signal events.}
\end{figure}
%

\subsubsection*{\boldmath Resonant ALP production}
At hadron colliders ALPs can be produced resonantly in gluon-gluon fusion. A gluon coupling implies the presence of di-jet final states, which are hard to distinguish from the background for masses $m_a < 1$ TeV. A more promising strategy is the search for di-photon events. Assuming non-vanishing couplings to photons and gluons, we show in Figure \ref{fig:gluonfusion} the sensitivity reach for the LHC, LHC$_{27}$ and FCC-hh in the $C_{GG}^{\rm eff}-C_{\gamma \gamma}^{\rm eff}$ plane. This reach is obtained by a rescaling of the constraint derived in the ATLAS analysis with $39.6\,$fb$^{-1}$ of data \cite{Aaboud:2017yyg}. The ALP production cross section is computed with \texttt{MadGraph5} \cite{Alwall:2014hca} and corrected for N$^3$LO corrections using the K factors $K_{gg} = 2.7$ at $m_a=200$ GeV,  $K_{gg} = 2.45$ at $m_a=500$ GeV and $K_{gg} = 2.35$ at $m_a=1$ TeV \cite{Ahmed:2016otz}.

\subsubsection*{\boldmath ALP production in exotic decays of $Z$ or Higgs bosons}

In analogy with the LHC specifications, we 
demand ALPs produced at $pp$ colliders and decaying into photons to decay inside the detector and before the electromagnetic colorimeter, $L_{\rm det} = 1.5\,$m, and for ALPs decaying into leptonic final states to decay before they reach the inner 
tracker, $L_{\rm det} = 2\,$cm. Our sensitivity reach is defined by requiring at least $100$ signal events. We use the reference cross sections $\sigma(gg 
\to h) = 146.6\,$pb \cite{Twiki} and $\sigma(pp \to Z) = 118.76 \,$nb  at $\sqrt{s} = 27\,$TeV, computed at NNLO \cite{Grazzini:2017mhc,Hamberg:1990np}. At $\sqrt{s} = 100\,$TeV, the relevant cross sections are $\sigma(gg \to h) = 802\,$pb and $
\sigma(pp \to Z) = 0.4\,\mu$b~\cite{Mangano:2016jyj}. 

%
\begin{figure}[t]
\begin{center}
\includegraphics[width=.75\textwidth]{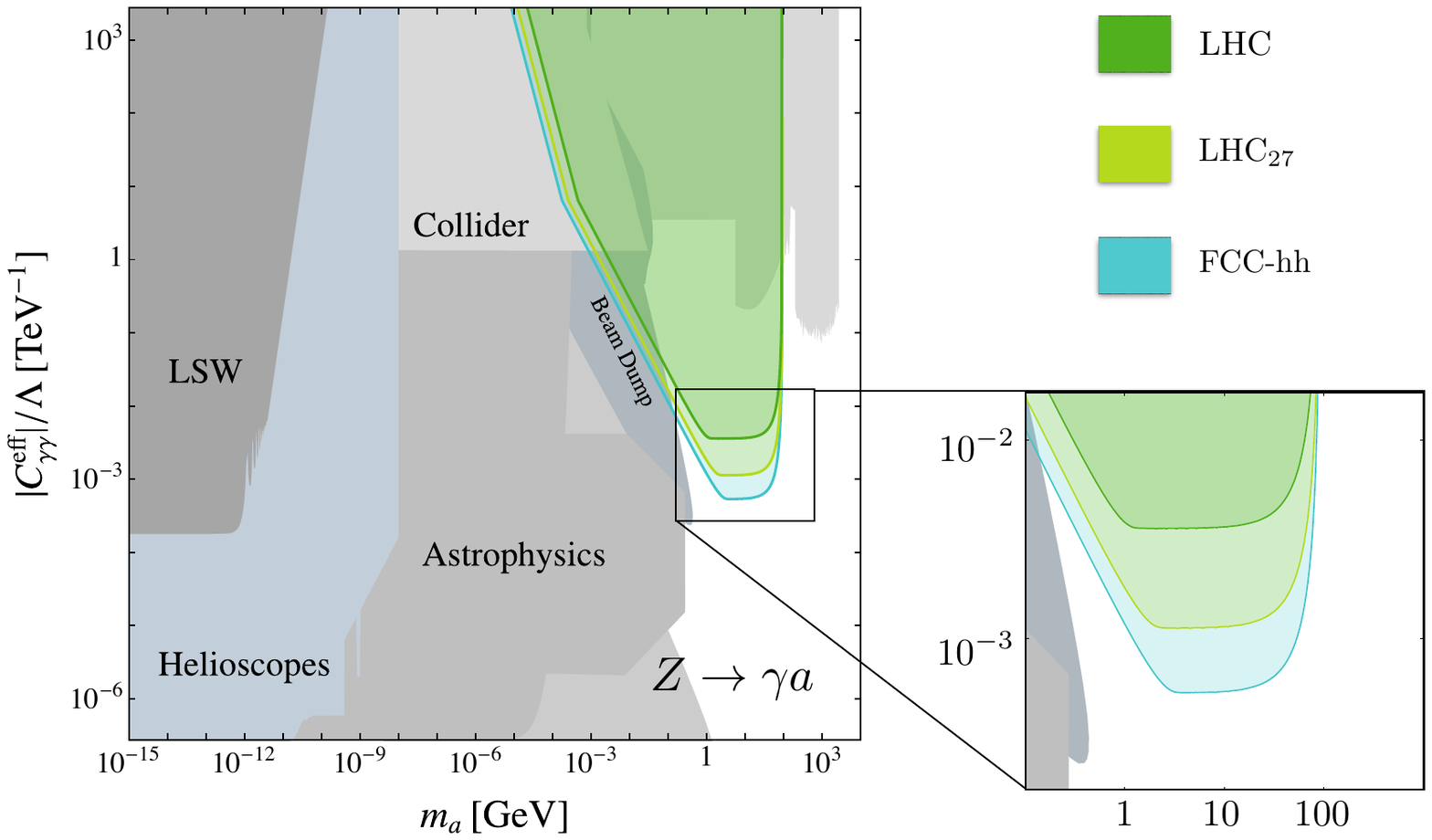}
\end{center}
\vspace{-3mm}
\caption{\label{fig:rareZdecayshh} Parameter regions which can be probed in the decay $Z \to \gamma a$ with $a \to \gamma \gamma$ at hadron colliders . The projected reach is coloured green (LHC), light green (HE-LHC) and turquoise (FCC-hh). We assume $\text{Br}(a\to\gamma\gamma)=1$. The sensitivity regions are based on 100 expected signal events.}
\end{figure}
%

In Figure~\ref{fig:rareZdecayshh} we show the reach of the LHC, the HE-LHC (LHC$_{27}$) and the FCC-hh in searches for $pp \to Z \to \gamma a\to 3\gamma$, assuming as before that $C_{WW}=0$ and $\text{Br}(a\to \gamma\gamma)=1$.
The reach of the HE-LHC extends beyond the reach of the LHC at $\sqrt{s}=14\,$TeV by a factor of about 3.2 assuming an integrated luminosity of $15\,$ab$^{-1}$. Colliders with $\sqrt{s}= 100\,$TeV and $20\,$fb$^{-1}$ can improve this reach by a factor of about 6.7 compared with the LHC. However, a high-luminosity run of an $e^+e^-$ collider on the $Z$-pole, as for example proposed for the FCC-ee, can probe the same couplings with even higher precision, as becomes clear by comparing the left upper panel of Figure~\ref{fig:eehabounds} with Figure~\ref{fig:rareZdecayshh}.

%
\begin{figure}[t]
\begin{center}
\includegraphics[width=0.9\textwidth]{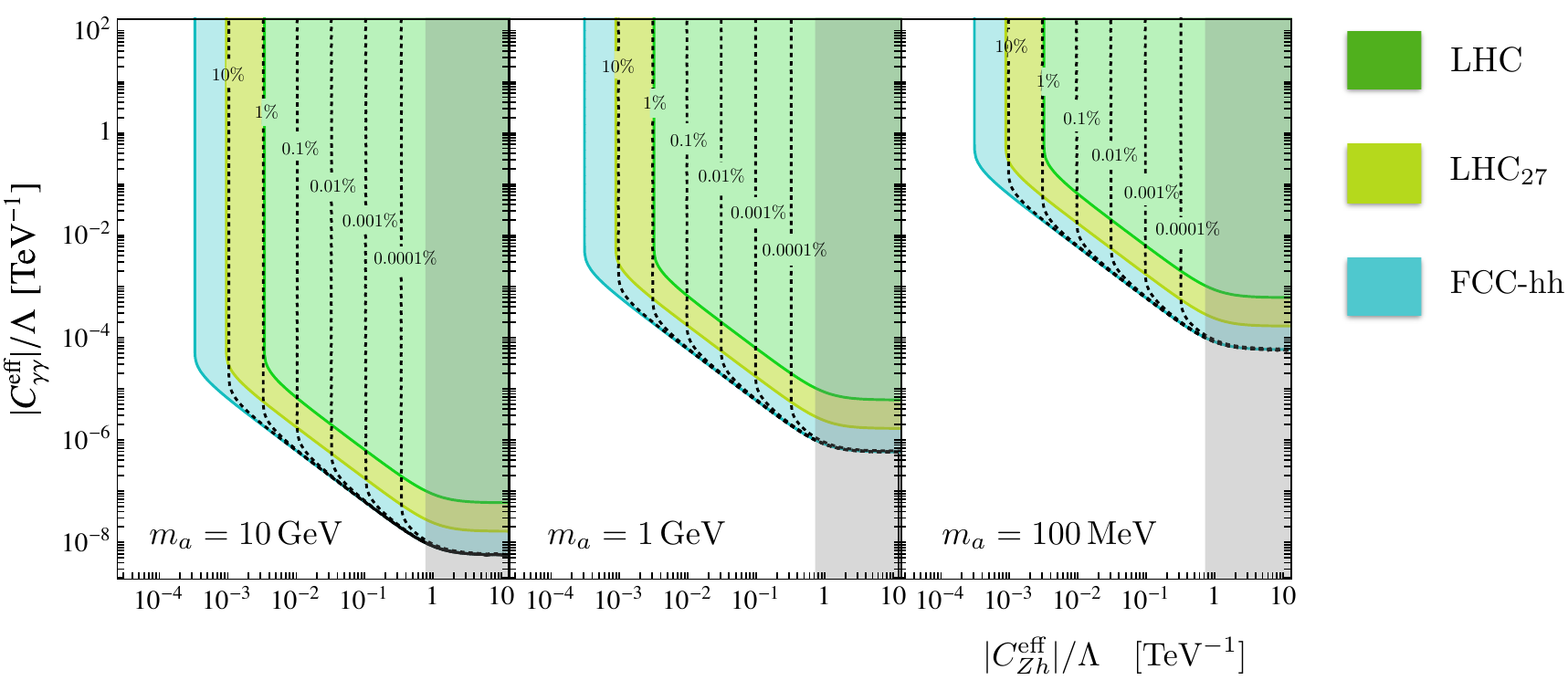}
\includegraphics[width=0.9\textwidth]{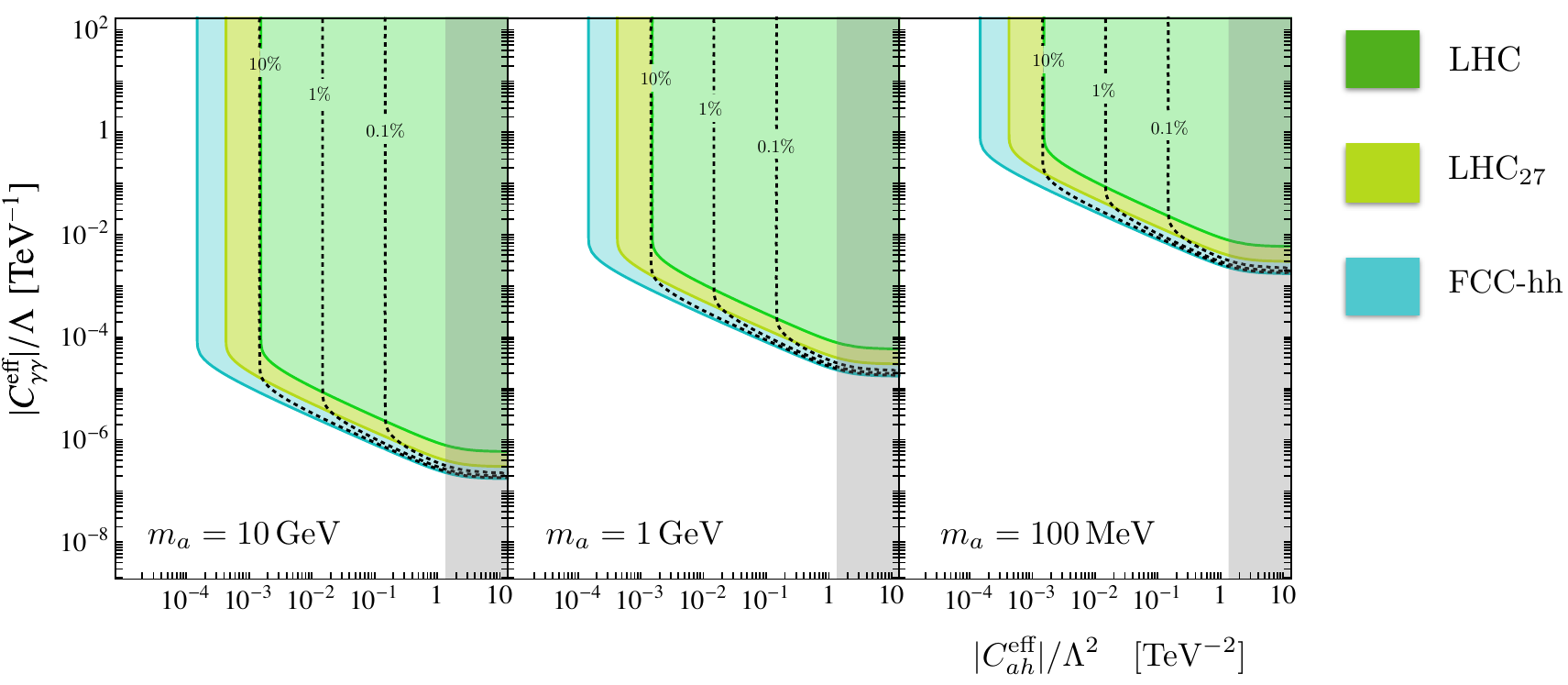}
\end{center}
\vspace{-0mm}
\caption{\label{fig:pphZa} Projected reach in searches for $h \to Za \to \ell^+\ell^-+2\gamma $ and $h \to aa \to 4\gamma $ decays with the LHC 
(green), HE-LHC (light green) and a $100\,$TeV collider (blue). The parameter region with the solid contours correspond to a branching ratio of $\text{Br}(a\to 
\gamma\gamma)=1$, and the contours showing the reach for smaller branching ratios are dotted. The sensitivity regions are based on 100 expected signal events.}
\end{figure}
%

%
\begin{figure}[t]
\begin{center}
\includegraphics[width=0.9\textwidth]{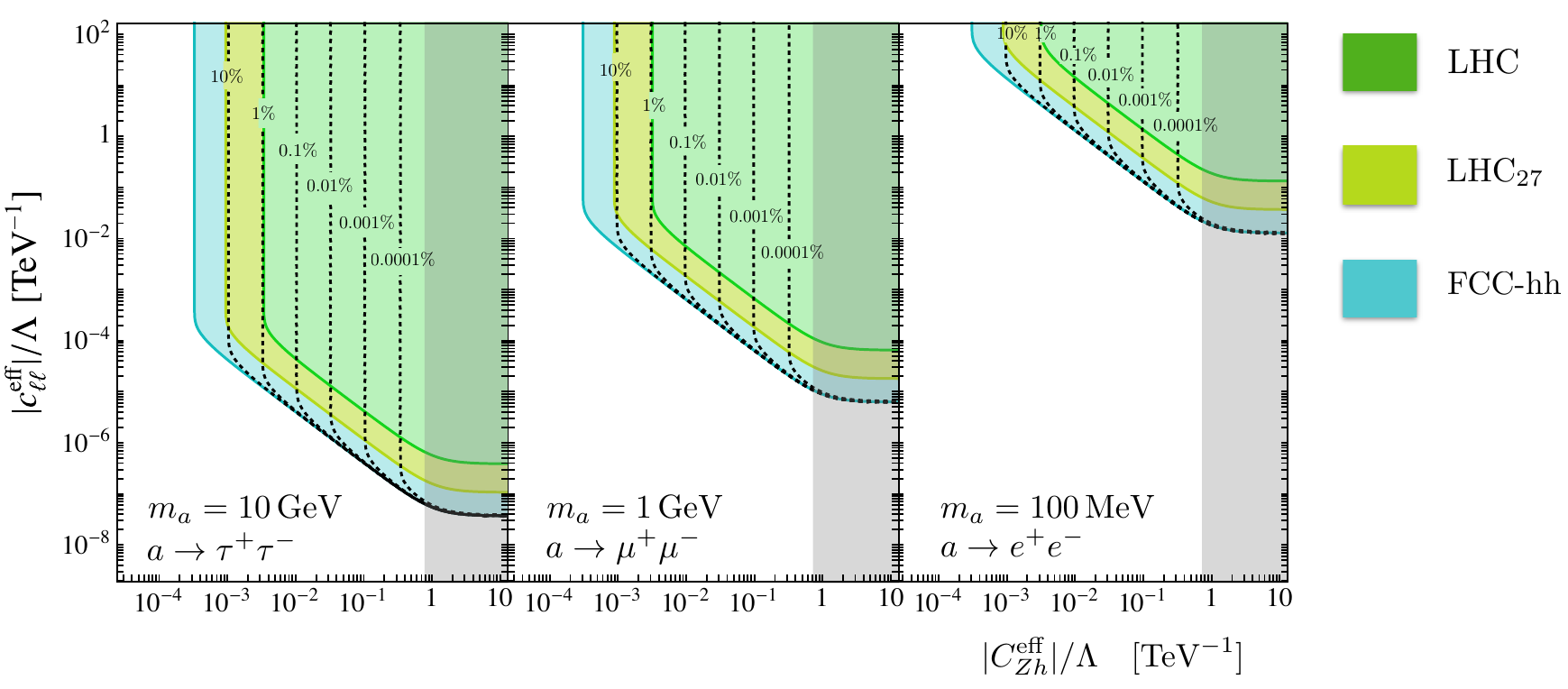}
\includegraphics[width=0.9\textwidth]{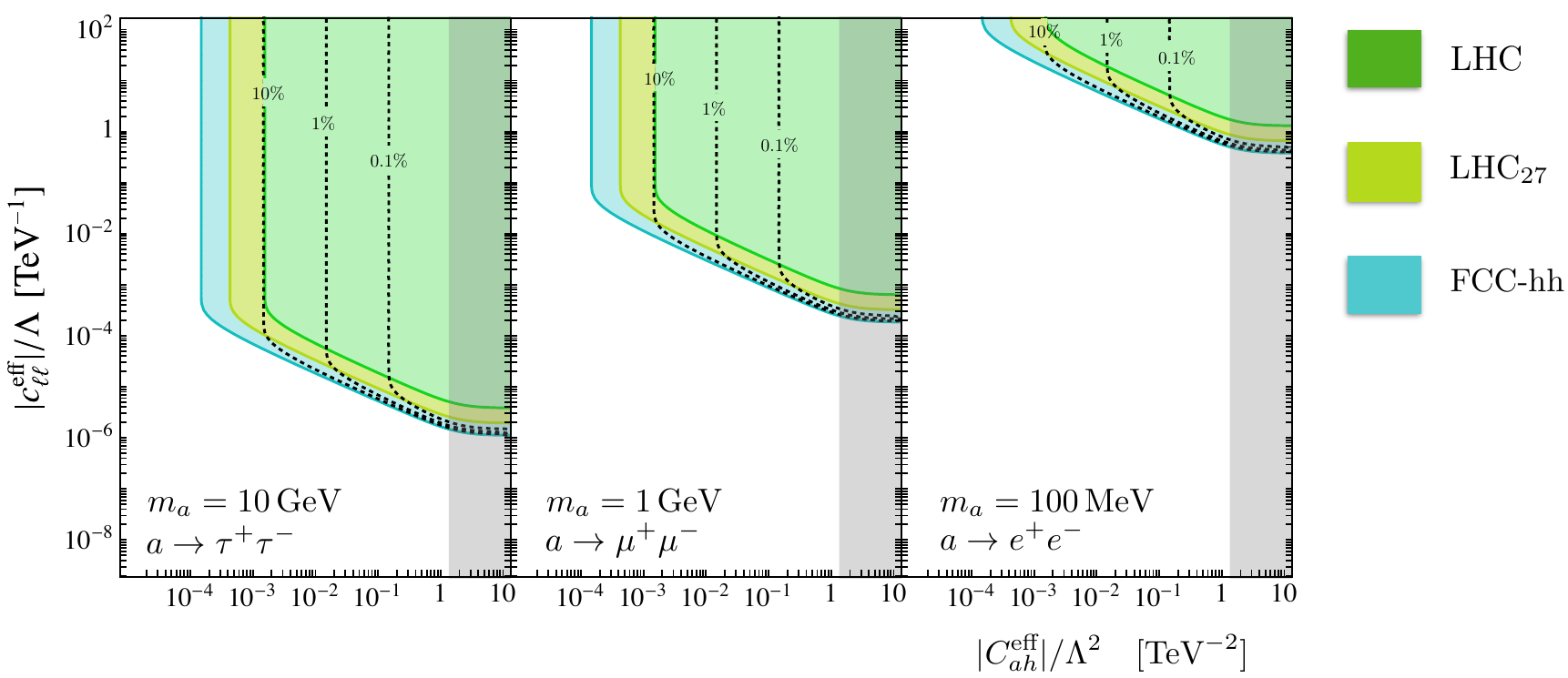}
\end{center}
\vspace{-3mm}
\caption{\label{fig:pphZalep} Projected reach in searches for $h \to Za \to \ell^+\ell^-+\ell^+\ell^- $ and $h \to aa \to 4\ell $ decays with the LHC 
(green), HE-LHC (light green) and a $100\,$TeV collider (blue). The parameter region with the solid contours correspond to a branching ratio of $\text{Br}(a\to 
\ell^+\ell^-)=1$, and the contours showing the reach for smaller branching ratios are dotted. The sensitivity regions are based on 100 expected signal events.}
\end{figure}
%

The situation is different for the case of exotic Higgs decays, because the Higgs production cross sections at hadron colliders with $\sqrt{s}=14-100\,$TeV are larger by orders of magnitude compared to the proposed future lepton colliders. In Figure~\ref{fig:pphZa}, we display the reach for observing 100 events at the LHC, HE-LHC and FCC-hh for searches for $pp\to h \to Za\to \ell^+\ell^-\gamma\gamma$ (upper panels) and  $pp\to h \to aa\to 4 \gamma$ (lower panels) for $m_a= 100\,$MeV, $1\,$GeV and $10\,$GeV and $\text{Br}(a\to \gamma\gamma)=1$. We further indicate the reach obtained in the case that $\text{Br}(a\to \gamma\gamma)<1$ by the dotted lines. Even though we rely on leptonic $Z$ decays with $\text{Br}(Z\to \ell^+\ell^-)=0.0673$ to account for the more challenging environment at hadron colliders, a future $100\,$TeV collider significantly improves beyond the projected reach in $C_{Zh}^{\rm eff}$ and $C_{ah}^{\rm eff}$ of the FCC-ee shown in Figure~\ref{fig:higgsdecaysL}. The sensitivity to $C_{\gamma\gamma}^\text{eff}$, however, is comparable between the FCC-ee and FCC-hh, and the projections for searches for $e^+e^-\to ha\to b\bar b \gamma\gamma$ at the second and third stage of CLIC even surpass the FCC-hh sensitivity in $C_{\gamma\gamma}^{\rm eff}$. 
For all considered ALP masses, the $h\to Z a$ decay could be observed at a $100\,$TeV collider for $\text{Br}(a\to \gamma\gamma)\gtrsim 10^{-6}$ and the $h\to a a$ decay could be fully reconstructed for $\text{Br}(a\to \gamma\gamma)\gtrsim 0.01$. \\
The results are similar for leptonic ALP decays. In Figure~\ref{fig:pphZalep} we show the reach in the $c_{\ell\ell}^\text{eff} - C_{Zh}^\text{eff}$ plane (upper row) and  $c_{\ell\ell}^\text{eff} - C_{ah}^\text{eff}$ plane (lower row). The results are again comparable with the projections for searches at future lepton colliders shown in Figure~\ref{fig:higgsdecaysLlep}.

%
\subsection{Searches for ALPs with macroscopic lifetime}\label{sec:MATHUSLA}
%

For small couplings and light ALPs produced in Higgs or $Z$ decays, the ALP decay vertex can be considerably displaced from the production vertex. For ALPs still decaying in the detector volume, this secondary vertex can be used to further suppress backgrounds. Very long-lived ALPs, which leave the detector before they decay, only leave a trace of missing energy. A detector further away from the interaction point can detect the decay products of these ALPs and reconstruct the ALP mass and direction. Recent proposals include the MATHUSLA large-volume surface detector \cite{Chou:2016lxi,Curtin:2018mvb} build above the ATLAS or CMS site at CERN, the Codex-B detector \cite{Gligorov:2017nwh} build in a shielded part of the LHC$b$ cavern, and a set of detectors called FASER \cite{Feng:2017uoz} build along the beam line, $\sim 150\,$m and $\sim 400\,$m from the interaction point of ATLAS or CMS. Since long lived ALPs are mostly produced in Higgs and $Z$ decays at the LHC, we will consider the reach of  the surface detector MATHUSLA for ALPs produced in the decays $Z\to \gamma a$, $h \to  Z a$ and $h \to aa$. We present projections for the sensitivity region for ALPs decaying into photons, muons and jets (gluons). Note that the possibility to detect photons with the MATHUSLA detector is an optional feature of the current design plan \cite{Curtin:2018mvb}.

%
\begin{figure}[t]\centering
\includegraphics[width=1\textwidth]{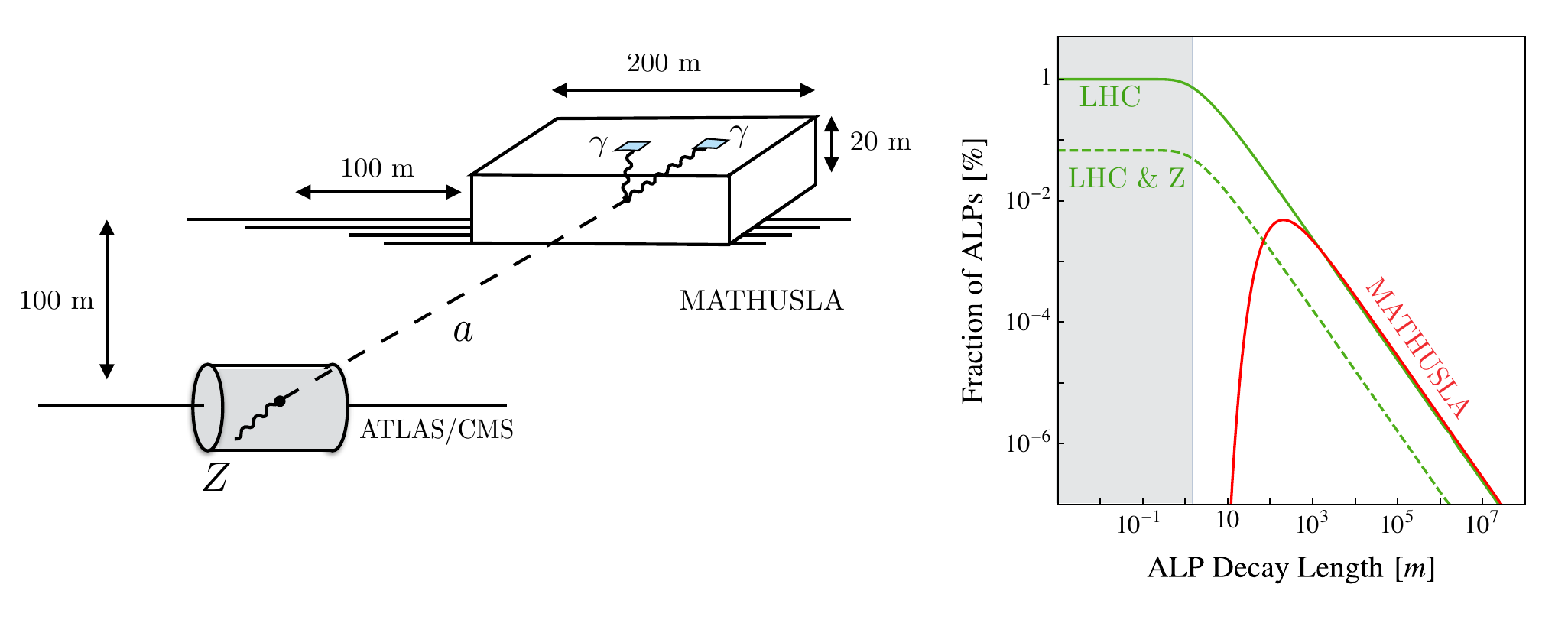}\\[0cm]
\caption{\label{fig:MATHUSLA} Left: Geometric setup of the MATHUSLA surface detector above the ATLAS/CMS cavern together with a sketch of the $pp\to h\to aZ$ process with a subsequent decay of the ALP in the MATHUSLA detector volume. Right: Total percentage of ALPs decaying within the ATLAS or CMS detector per ALPs produced in the Higgs decay $h\to aZ$ (green), fraction of ALPs produced decaying in ATLAS/CMS together with a leptonically decaying $Z$ (dashed green), and the percentage of ALPs decaying within the MATHUSLA detector volume (red). The grey area shows the distance between the interaction point and the electromagnetic calorimeter.}
\end{figure}
%

For MATHUSLA, it is impossible to detect both final state particles in $h\to Za $ and $Z\to  \gamma a$ decays and highly unlikely to see both ALPs from $h\to aa$ 
decays in the decay volume. However, because of the much lower background, single ALPs can be detected irrespective of their origin. The fraction of ALPs decaying in the MATHUSLA detector is then given by 
\begin{align}\label{eq:faM}
f^a_\text{M}=\int_{\Omega_\text{M}}  d\Omega \,\bigg(\frac{1}{\sigma}\,\frac{d\sigma}{d\Omega}\bigg)\,
\Big[e^{-r_\text{in}(\Omega)/L_a}-e^{-r_\text{out}(\Omega)/L_a}\Big]\,,
\end{align}
where $\Omega_\text{M}$ describes the area in solid angle covered by the MATHUSLA detector, $d\sigma/d\Omega$  denotes the differential cross section for ALPs produced in the decay of a $Z$ or Higgs boson in the laboratory frame, and $L_a=p_a/(\Gamma_a m_a)$, where $p_a$ is the ALP momentum in that frame. At fixed solid angle, the radii $r_\text{in}$ and $r_\text{out}$ denote the distances between the interaction point and the intersections of the ALP line of flight with the MATHUSLA detector. The MATHUSLA detector with a volume of $20\,\text{m} \times 200\,\text{m} \times 200\,\text{m}$ will be placed $100\,$m above the beam line and $100\,$m shifted from the interaction point along the beam line and has a considerably smaller coverage in solid angle: approximately $5\%$ at MATHUSLA compared to $100\%$ at ATLAS and CMS. Nevertheless, as Figure~\ref{fig:MATHUSLA} shows, for long-lived ALPs, the number of ALPs decaying in the MATHUSLA volume is comparable to the number of ALPs decaying within a radius of $1.5\,$m from the interaction point. However, for ALPs with masses $m_a>1\,$GeV backgrounds at MATHUSLA are negligible, whereas for example for $h \to Z a $ decays the $Z$ boson needs to be reconstructed and more events are required to distinguish the signal from the background. As in Section \ref{sec:hadcol}, we therefore demand at least 100 events with leptonically decaying $Z$ boson to determine the LHC reach, and at least 4 reconstructed ALP decays to determine the reach of MATHUSLA.
In the left panel of Figure~\ref{fig:MATHUSLA} we illustrate the geometry of the proposed MATHUSLA experiment. The right panel shows the percentage of ALPs produced via $pp\to h\to Z a$ that decay before reaching the electromagnetic calorimeter (green), the percentage of ALPs decaying within the detector together with a leptonically decaying $Z$-boson (dashed green), and the percentage of ALPs decaying within the MATHUSLA detector volume (red) as a function of the ALP decay length. Taking into account the additional relative factor of $\sim1/20$ between the number of events we expect to determine the reach of LHC and MATHUSLA, the MATHUSLA detector performs significantly better than the LHC for ALPs with a decay length exceeding $100\,$m.

%
%
\begin{figure}[t]\centering
\includegraphics[width=.95\textwidth]{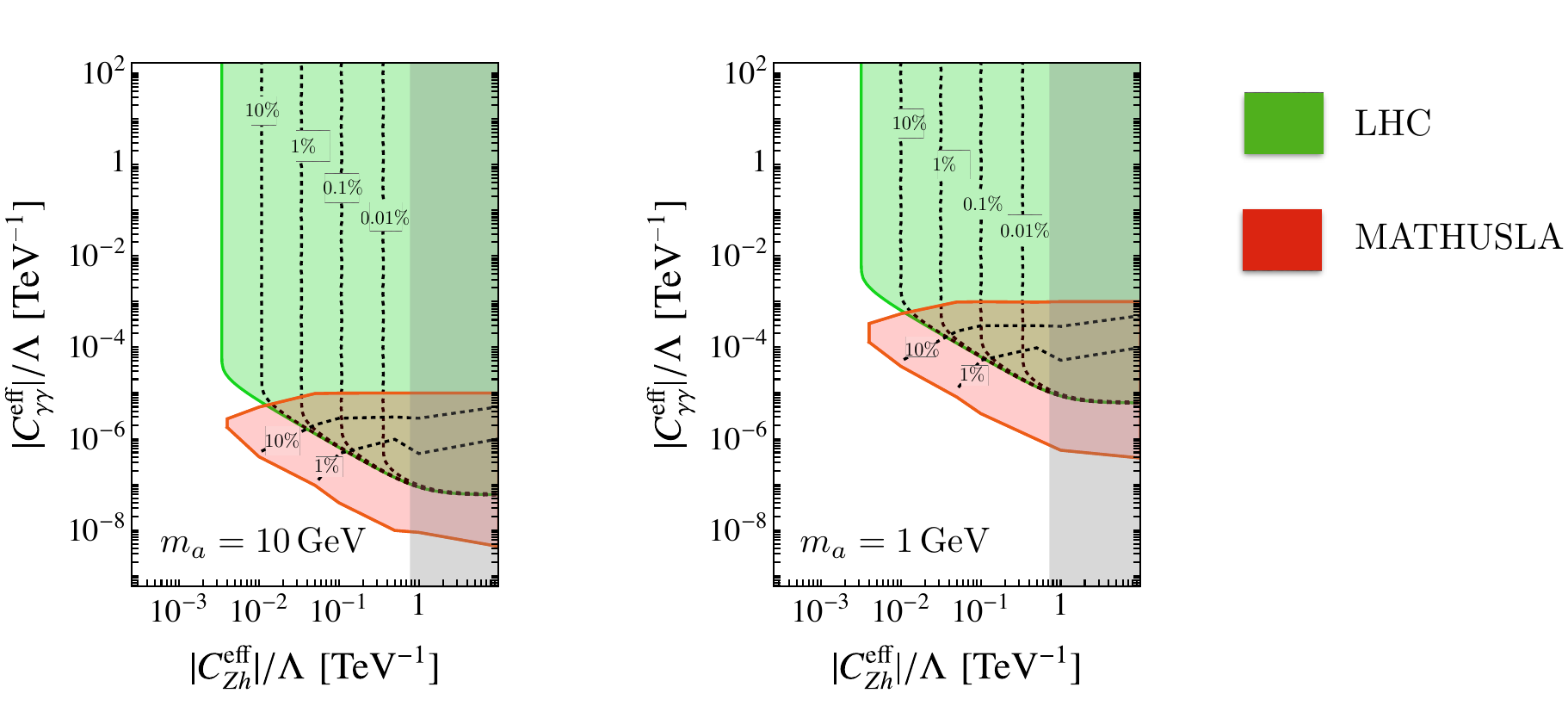}\\[0cm]
\includegraphics[width=.95\textwidth]{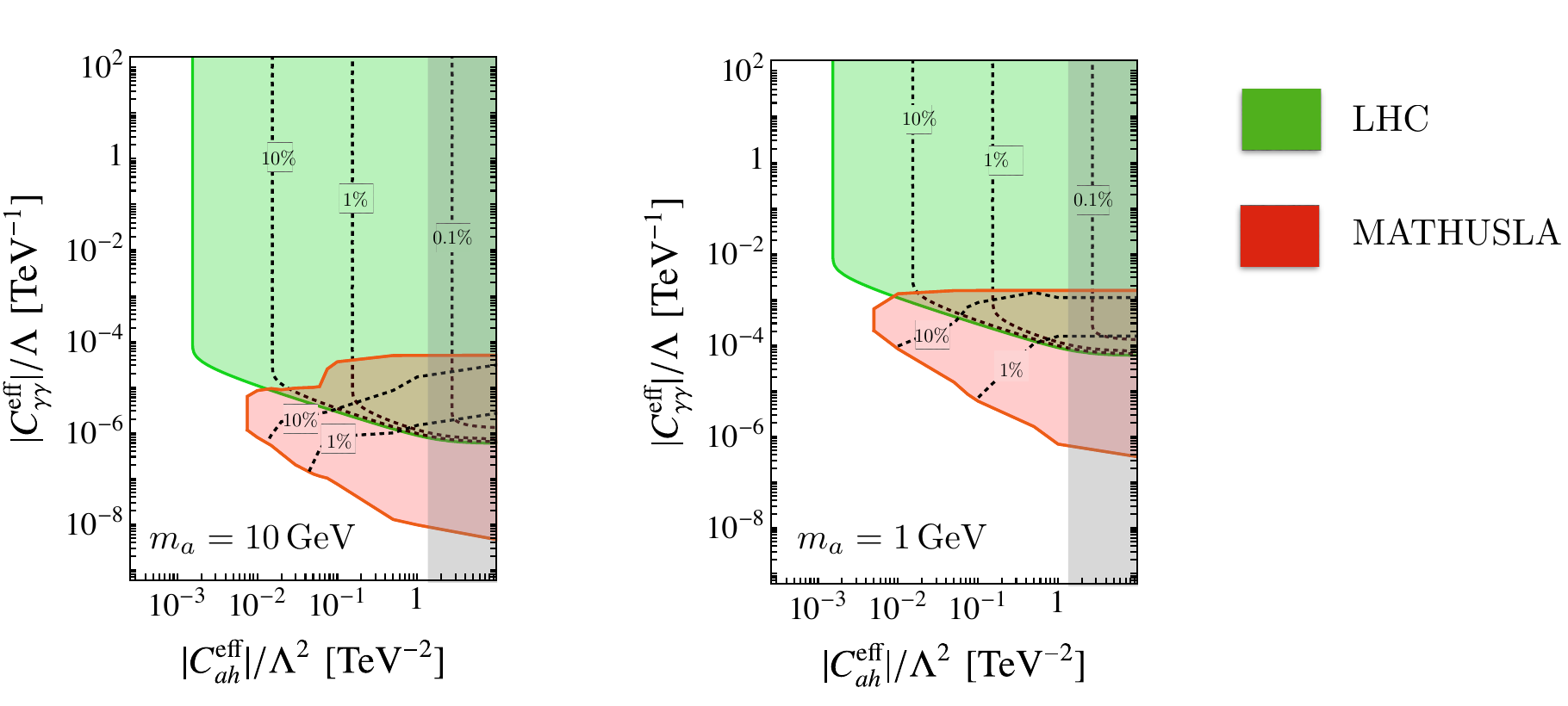}
\caption{\label{fig:hZa} Projected reach in searches for $h \to Za \to \ell^+\ell^-+2\gamma$ (top) and $h \to aa \to4\gamma$ (bottom) decays at the LHC 
(green) and MATHUSLA (red) with $\sqrt{s}=14\,$TeV center-of-mass energy and $3\,$ab$^{-1}$ integrated luminosity. The parameter region with solid 
contours correspond to a branching ratio of $\text{Br}(a\to \gamma\gamma)=1$, and contours showing the reach for smaller branching ratios are dotted. The sensitivity regions are based on 4 (MATHUSLA) and 100 (LHC) expected signal events, respectively .}
\end{figure}
%

Using \eqref{eq:faM}, we can define the corresponding effective branching ratios for ALP decays in MATHUSLA in analogy with \eqref{eq:LHCZag},
\begin{align}
\text{Br}(h\to Za\to Z \gamma\gamma)\big\vert_\text{eff}^\text{M}&=\text{Br}(h\to Za)\,\text{Br}(a\to \gamma\gamma)f_\text{M}^a\,\label{eq:MATHhZa}\,,\\
\text{Br}(h\to aa\to a \gamma\gamma)\big\vert_\text{eff}^\text{M}&=2\text{Br}(h\to aa)\,\text{Br}(a\to \gamma\gamma)f_\text{M}^a\,\label{eq:MATHhaa}\,,\\
\text{Br}(Z\to \gamma a\to 3 \gamma)\big\vert_\text{eff}^\text{M}&=\text{Br}(Z\to \gamma a)\,\text{Br}(a\to \gamma\gamma)f_\text{M}^a\, \label{eq:MATHZag}.
\end{align}
The expressions for ALP decays into leptons are analogous with the ALP decay into photons with $\text{Br}(a\to\gamma\gamma)$ replaced by $\text{Br}(a\to\ell^+\ell^-)$. In order to fully capture the geometric acceptance of the MATHUSLA detector, we use \texttt{MadGraph5} to simulate the signal events at parton level and the code provided by the MATHUSLA working group to compute the acceptance \cite{Curtin:2018mvb}.

We illustrate the reach of the LHC and the MATHUSLA detector for discovering ALPs decaying into photons from $h \to Za$ (upper panels) and $h \to a a$ (lower panels) decays 
in Figure~\ref{fig:hZa}. For the green region with solid contours, the LHC would see 100 events with a 
branching ratio of $\text{Br}(a\to \gamma\gamma)=1$. For smaller branching ratios, larger couplings $|C^\text{eff}_{hZ}|$ and $|C_{ah}^{\rm eff}|$ are required to obtain the 
same number of events. Dotted lines show the lower limit for $\text{Br}(a\to \gamma\gamma)=0.1$ and $\text{Br}(a\to \gamma\gamma)=0.01$. The red region with solid contours shows the parameter space for which 4 ALP decays are expected within the MATHUSLA detector 
volume for $\text{Br}(a\to \gamma\gamma)=1$. Smaller branching ratios with constant partial width for ALP decays into photons imply a larger total decay width of 
the ALP and therefore smaller decay lengths. For $\text{Br}(a\to \gamma\gamma)=0.1$ and $\text{Br}(a\to \gamma\gamma)=0.01$, MATHUSLA therefore looses sensitivity for  larger values of $|C_{\gamma\gamma}^{\rm eff}|/\Lambda$. In the case of $h \to a a$ decays, MATHUSLA 
will be able to probe smaller branching ratios than ATLAS and CMS. This underlines the complementarity between searches for prompt decays with ATLAS/CMS and searches for displaced ALP decays with MATHUSLA.  
We stress that a discovery of a resonance with MATHUSLA alone cannot be used to determine the production mode of the ALP. However, one can use the reconstructed mass of the ALP and the number of observed events to guide future searches at the LHC, for example searches 
for invisible ALPs in the final state. 

%
\begin{figure}\centering
\includegraphics[width=.95\textwidth]{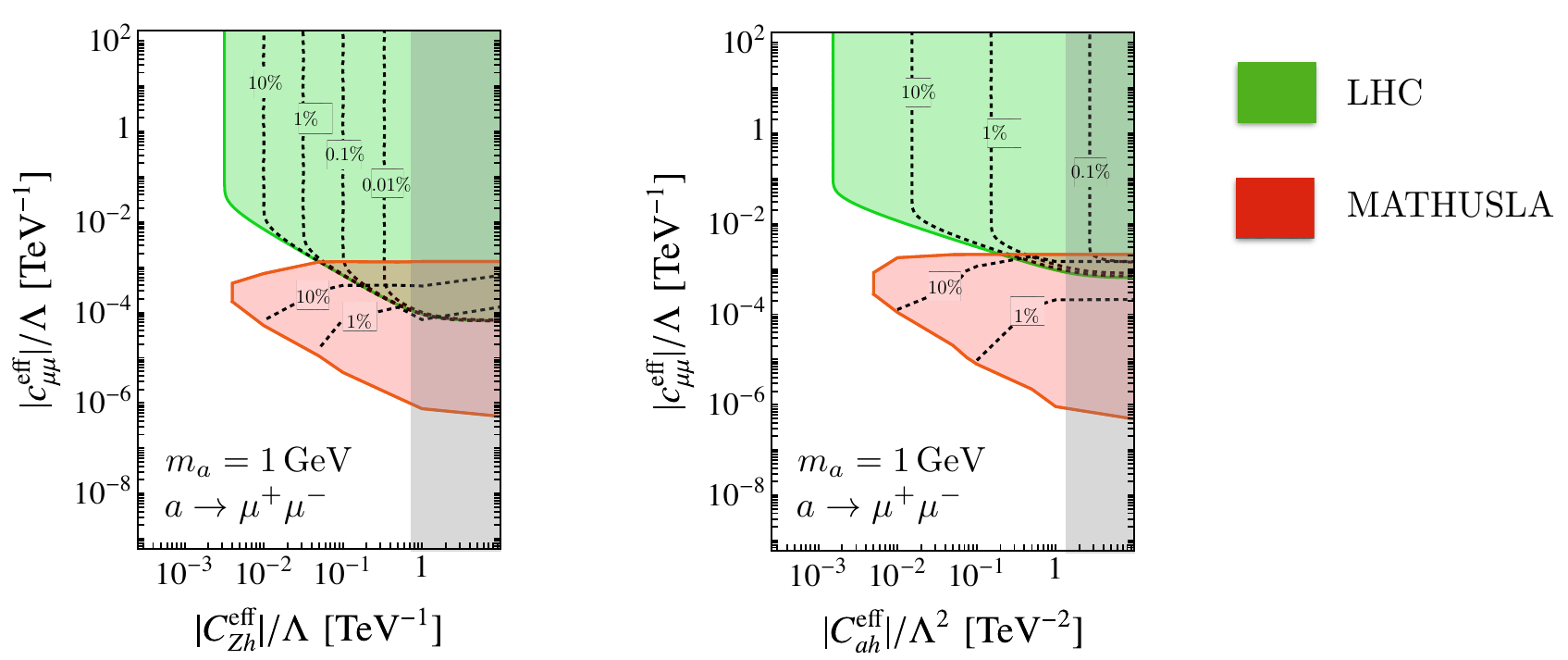}
\caption{\label{fig:muons} Projected reach in searches for $h \to Za \to \ell^+\ell^-+\mu^+\mu^- $ (left) and $h \to aa \to \mu^+\mu^-+\mu^+\mu^- $ (right) decays 
with ATLAS/CMS (green) and MATHUSLA (red) with $\sqrt{s}=14\,$TeV center-of-mass energy and $3\,$ab$^{-1}$ integrated luminosity. The parameter region 
with solid contours correspond to a branching ratio of $\text{Br}(a\to \mu^+\mu^-)=1$, and contours showing the reach for smaller branching ratios are 
dotted. The sensitivity regions are based on 4 (MATHUSLA) and 100 (LHC) expected signal events, respectively.}
\end{figure}
%
%
\begin{figure}
\centering
\includegraphics[width=.8\textwidth]{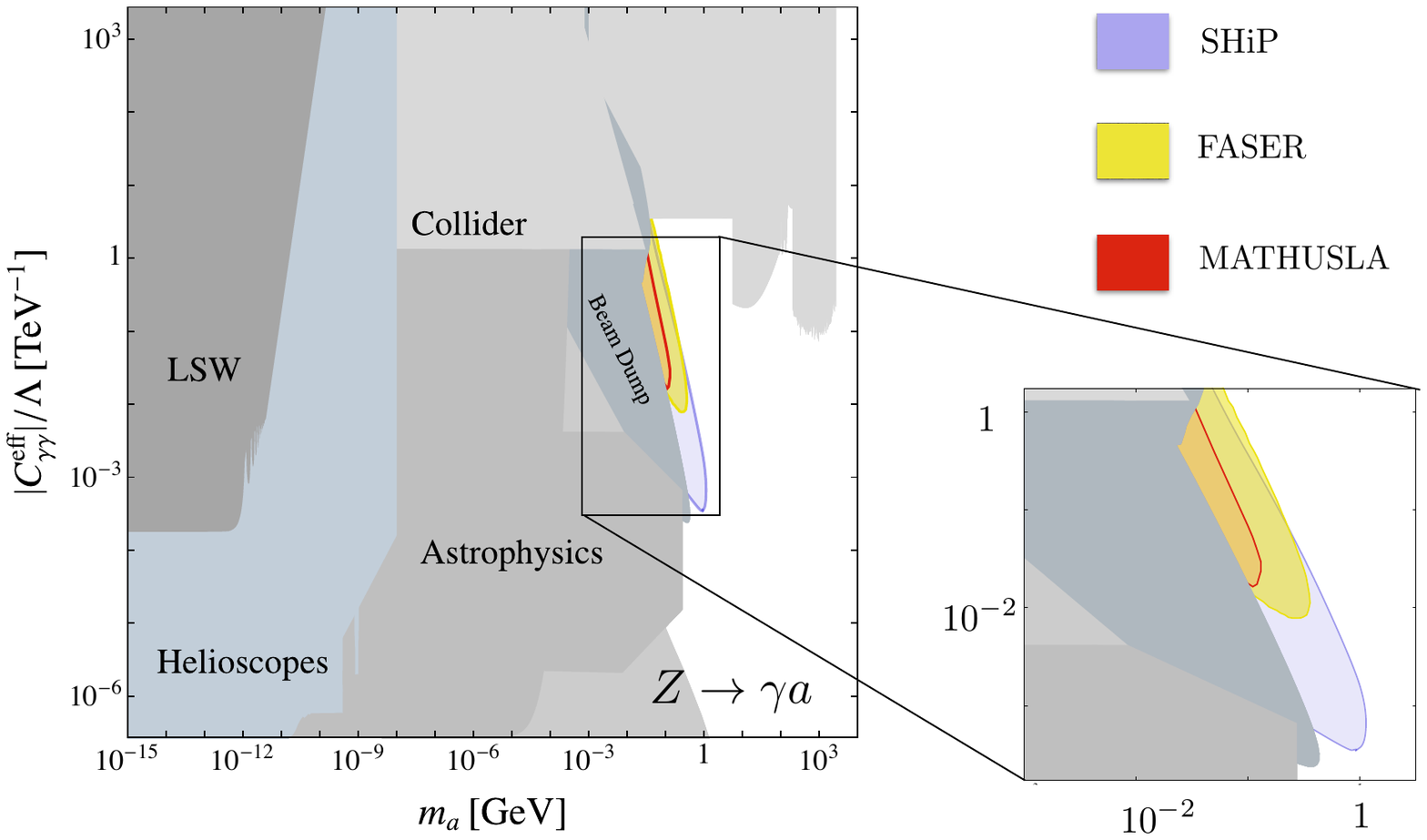}
\caption{\label{fig:ZAG} Projected reach in searches for $Z \to \gamma a \to 3\gamma$ with 
MATHUSLA  for $\sqrt{s}=14\,$TeV center-of-mass energy, $3\,$ab$^{-1}$ integrated luminosity and  
$\text{Br}(a\to \gamma\gamma)=1$, together with the expected sensitivity of FASER taken from 
\cite{Feng:2018pew} and SHiP \cite{Alekhin:2015byh}. The sensitivity regions are based on 4 (MATHUSLA) and 100 (LHC) expected signal events, respectively.}
\end{figure}
%

In Figure~\ref{fig:muons}, we show the reach of $h \to Za$ and $h \to aa$ for ALPs decaying into muons. Since at least approximate lepton-flavor universality is 
expected for the couplings of the ALP, the muon decay mode is particularly well motivated for $2m_\mu < m_a < 2m_\tau$. Also here,
MATHUSLA can probe much smaller couplings $|c_{\mu\mu}^{\rm eff}|$ than the LHC.

In the case of $Z\to \gamma a $ decays, we show the reach of MATHUSLA in the $m_a - |C_{\gamma\gamma}^{\rm eff}|/\Lambda$ plane in Figure~\ref{fig:ZAG}, again assuming $C_{WW}=0$. In principle, for non-vanishing $C_{\gamma Z}$, searches for exotic $Z$ decays with MATHUSLA compete with the reach of future beam-dump experiments such as ShiP \cite{Alekhin:2015byh}. However for light ALPs, the reach shown in Figure~\ref{fig:ZAG} is probably overestimated. Whether the MATHUSLA detector will be able to resolve photon pairs for $m_a< 1\,$GeV will depend on the angular resolution of the final detector proposal. Interestingly, FASER can take advantage of the large Primakoff cross section for photons producing ALPs through interaction with the detector material ($\gamma N \to a N$) in the forward region to set limits on $C_{\gamma\gamma}^{\rm eff}$ independently \cite{Feng:2018pew}. The corresponding projected sensitivity reach of FASER is slightly better than that of MATHUSLA. 

%
\begin{figure}[t]
\begin{center}
\includegraphics[width=.99\textwidth]{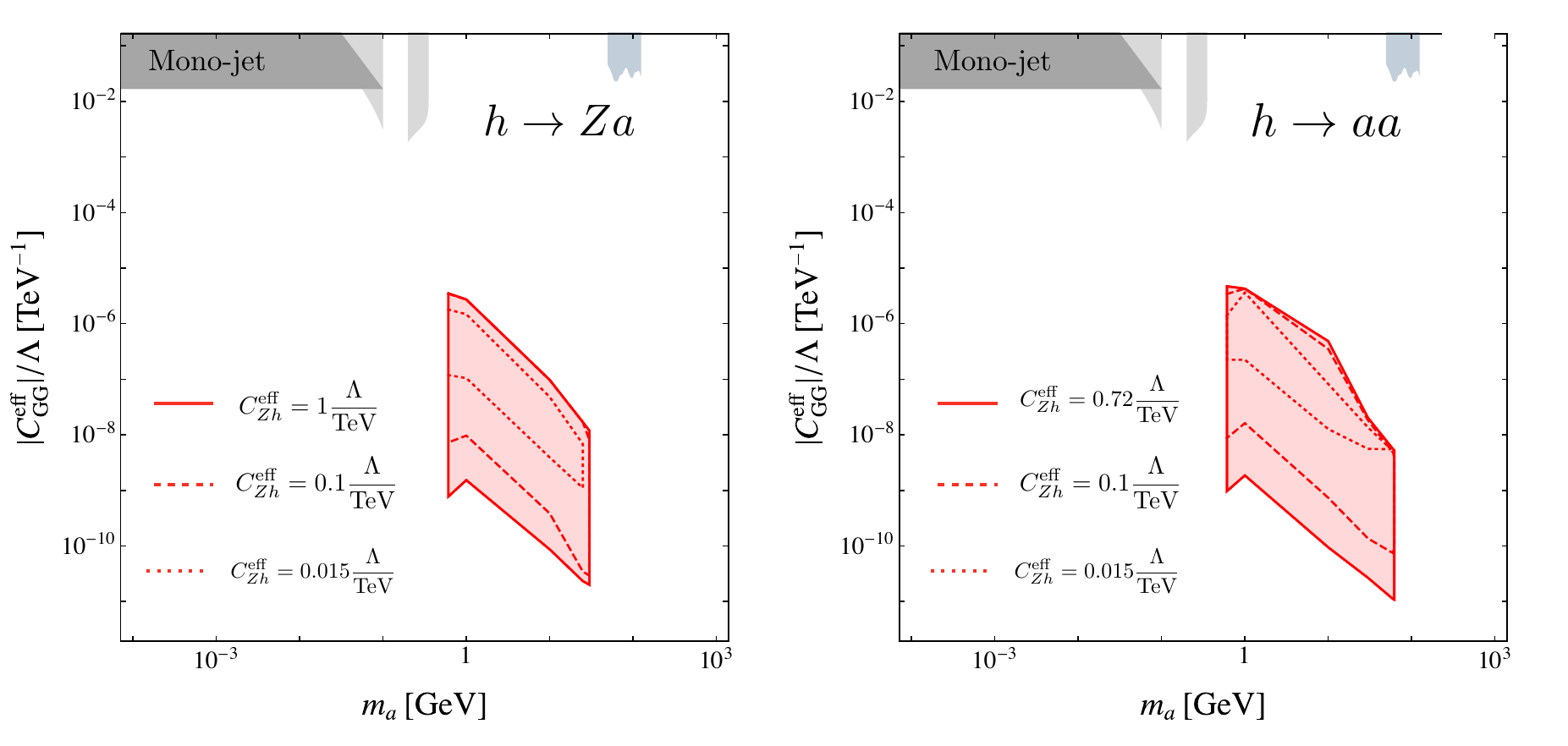}
\end{center}
\vspace{-3mm}
\caption{\label{fig:Mathuslagluonlimits} Projected exclusion contours for searches for $pp\to h \to Z a $  (left) and $pp\to h \to aa $ (right) with the subsequent ALP decay $a \to gg$ and $\text{Br}(a\to gg)=1$ with the MATHUSLA detector. The different contours correspond to different values of $C_{Zh}^\text{eff}$ and $C_{ah}^\text{eff}$. The sensitivity regions are based on 4 expected signal events, respectively.}
\end{figure}
%

A unique strength of surface detectors is the possibility to constrain hadronic ALP decays, whereas light ALPs ($m_a< 500\,$GeV) decaying into jets are hard to detect at the LHC because of the large QCD background. For ALPs produced in gluon fusion or through ALP-quark couplings, a sizeable production cross section corresponds to couplings too large to produce any signal in the MATHUSLA detector. ALPs produced in resonant Higgs or $Z$ decays can be detected in MATHUSLA by reconstructing di-jet (or multi-jet) events. Particularly well motivated are ALPs with only couplings to gluons, because in models addressing the strong CP problem the ALP-gluon coupling is the only ALP coupling that cannot be avoided.
We show the parameter space for which at least four $a\to jj$ events are expected within the MATHUSLA volume in the $m_a-C_{GG}^\text{eff}$ plane in Figure~\ref{fig:Mathuslagluonlimits} for different values of $C_{Zh}^\text{eff}$ (left) and $C_{ah}^\text{eff}$ (right). The expected minimal mass resolution of the MATHUSLA detector for ALPs in Higgs decays is of the order of $m_a\approx 100\,$MeV, assuming a spatial resolution of $1\,$cm. In Figure~\ref{fig:Mathuslagluonlimits} the lowest ALP mass is $m_a= 600$ MeV. \footnote{Note that for ALP masses below $m_a=1\,$GeV the ALP-gluon coupling $C_\text{GG}^{\rm eff}$ induces a sizeable photon coupling through ALP-meson mixing, leading to additional constraints.}

%
\section{Conclusions}\label{sec:conclusions}
%

Any ultraviolet completion of the SM in which an approximate global symmetry is broken gives rise to pseudo-Nambu-Goldstone bosons, which are light with respect to the symmetry breaking scale $m_a  \ll \Lambda$. The discovery of such ALPs at the LHC or future colliders could therefore be the first sign of a whole sector of new physics, and measuring its properties could reveal important hints about the UV theory. 

We consider the most general effective Lagrangian including the leading operators in the $1/\Lambda$ expansion that couple the ALP to SM particles. Whereas couplings to SM fermions and gauge bosons can arise at mass dimension-5, the Higgs portal only arises at dimension-6. 
We derive projections for the most promising ALP search channels for the LHC, its potential future high-energy upgrade, as well as a variety of possible future high-energy hadron and lepton colliders.

At lepton colliders, ALP production in association with a photon, a $Z$ boson or a Higgs boson provide the dominant production processes, provided the ALP couplings to either hypercharge, $SU(2)_L$ gauge bosons or to the Higgs boson are present in the Lagrangian. 
Even if only ALP-fermion couplings are present at tree-level, ALP couplings to gauge bosons  are generated at one-loop order through the anomaly equation. We point out that a high-luminosity run at the $Z$ pole would significantly increase the sensitivity to ALPs produced in $e^+e^-\to \gamma a$ with subsequent decays $a\to \gamma\gamma$ or $a\to \ell^+\ell^-$. This favours the FCC-ee proposal over CLIC in these particular searches, whereas CLIC, operating at $\sqrt{s}=1.5\,$TeV or $\sqrt{s}=3\,$TeV, can discover significantly heavier ALPs. 

At hadron colliders ALPs can be produced copiously in gluon-fusion and via exotic $Z \to a \gamma$, $h \to a Z$ and $h \to a a$ decays. Searches for exotic $Z$ decays at a future 100 TeV collider are less sensitive to ALP-photon couplings than a high-luminosity run of the FCC-ee at the $Z$ pole. For the exotic Higgs decays $h \to Z a $ and $h \to a a$ already the LHC at $\sqrt{s}=14\,$TeV and $3\,$ab$^{-1}$ provides a better reach compared to future $e^+e^-$ colliders in the corresponding Wilson coefficients $C_{ah}^{\rm eff}$ and $C_{Zh}^{\rm eff}$.  The sensitivity of a future $100\,$TeV collider in both $C_{Zh}^{\rm eff}$ and $C_{ah}^{\rm eff}$ is about an order of magnitude larger than at the LHC, and about a factor of $3$ in the coefficients $C_{\gamma\gamma}^{\rm eff}$ (for $a\to \gamma\gamma$) and $c_{\ell\ell}^{\rm eff}$ (for $a\to \ell^+\ell^-$). 

A future dedicated detector searching for long-lived particles at the LHC,  such as MATHUSLA, FASER or Codex-B could provide sensitivity for even smaller ALP couplings to photons, charged leptons or jets. MATHUSLA has unique capabilities to search for long-lived ALPs with a mean decay length of $100\,$m and more, corresponding to couplings 2-3 orders of magnitude smaller than the ones that can be probed with ATLAS and CMS. Such ALPs cannot be produced resonantly with a significant cross section, but large numbers of ALPs with small widths can be produced in exotic decays of Higgs or $Z$ bosons. The main backgrounds at MATHUSLA are cosmic rays, allowing for a cleaner environment for observing ALPs in the $\mathcal{O}(1)-\mathcal{O}(10)\,$GeV range. This is particularly powerful for hadronically decaying ALPs, where MATHUSLA can overcome the large QCD background at the LHC and thus provide the opportunity to constrain light ALPs decaying into jets, which are otherwise difficult due to the large QCD background at hadron colliders. 

Long-lived ALPs or ALPs that couple to dark matter \cite{Nomura:2008ru} can also be searched for by cutting on missing energy. The focus of this paper is on ALPs that can be reconstructed from their decay products, but projections for searches for missing energy signatures at the LHC with $3000$ fb$^{-1}$ have been presented in \cite{Brivio:2017ije}, and for a future ILC and TLEP with a center of mass energy of $240$ GeV and 1 TeV, respectively in \cite{Mimasu:2014nea}. Since we demand the ALPs to decay within the detector for our projections, the part of the parameter space to which missing energy searches are sensitive is largely complementary to the parameter space for which ALPs can be discovered by the searches discussed in this paper.

\section*{Acknowledgements}
We thank Philipp Roloff for discussion of the CLIC detector design. The research reported here has been supported by the Cluster of Excellence {\em Precision Physics, Fundamental Interactions and Structure of Matter\/} (PRISMA -- EXC 1098), and grant 05H12UME of the German Federal Ministry for Education and Research (BMBF).

\addcontentsline{toc}{section}{References}
\bibliographystyle{JHEP}
\bibliography{references}

\end{document}